%% file: paper.tex
\documentclass{article} 
\usepackage{iclr2027_conference,times}
\iclrfinalcopy

\usepackage{microtype}
\usepackage{graphicx}
\usepackage{subfigure}
\usepackage{booktabs} 

\usepackage{hyperref}

\usepackage{amsmath}
\usepackage{amssymb}
\usepackage{mathtools}
\usepackage{amsthm}

\usepackage[capitalize,noabbrev]{cleveref}

\theoremstyle{plain}
\newtheorem{theorem}{Theorem}[section]

\newtheorem{lemma}[theorem]{Lemma}

\theoremstyle{definition}

\theoremstyle{remark}

\usepackage[textsize=tiny]{todonotes}

\usepackage{graphicx}
\usepackage{amsmath,amsfonts}
\usepackage{amsthm}
\usepackage{balance}
\usepackage{xcolor}
\usepackage{fancyvrb}
\usepackage{algorithm}
\usepackage{algorithmic}
\usepackage{comment}
\usepackage{subcaption}
\usepackage{xspace}
\usepackage{paralist}
\usepackage{url}
\usepackage{pgfplots}
\usepackage{pgfplotstable}
\pgfplotsset{compat=1.18}
\usepackage{tikz}

\usepackage{bm}
\usepackage{wrapfig}
\usepackage{inputenc}  
\usepackage{lipsum}  
\usepackage{caption}
\usepackage{indentfirst}
\usepackage[htt]{hyphenat} 
\usepackage[hang,flushmargin]{footmisc}  
\usepackage{mathtools}
\usepackage{bold-extra}
\usepackage[all]{nowidow}
\usepackage{booktabs,multirow}
\usepackage{hyperref}
\usepackage{placeins}
\usepackage{pgfplots}
\usepackage{hyperref}
\usepackage{pgfplotstable}

\usepackage{MnSymbol}
\usepackage{pgfplots}
\usepackage{placeins} 

\newenvironment{sidealgorithm}[2]{%
    \begin{minipage}[t]{0.49\linewidth}
    \vspace{0pt}
    \refstepcounter{algorithm}\label{#2}%
    \hrule
    \vspace{2pt}
    {\raggedright\textbf{Algorithm \thealgorithm}~#1\par}
    \vspace{2pt}
    \hrule
    \vspace{3pt}
    \scriptsize
}{%
    \vspace{2pt}
    \hrule
    \end{minipage}
}

\input{savespace}

\begin{document}

\title{\textsf{SilentWood}: Efficient Private Inference Over Gradient Boosting Decision Forests}

\author{%
\parbox{\dimexpr\textwidth-2\tabcolsep\relax}{%
\raggedright
\normalfont
\textbf{%
Ronny Ko\textsuperscript{1},
Abdelkarim Kati\textsuperscript{2},
Robin Geelen\textsuperscript{3},
Rasoul Mahdavi\textsuperscript{2},
Byoungwoo Yoon\textsuperscript{1},
Jongho Shin\textsuperscript{1},
Zhiqiang Lin\textsuperscript{4},
Makoto Onizuka\textsuperscript{5},
Florian Kerschbaum\textsuperscript{2}
}
\\
\vspace{0.8em}
{\small
\textsuperscript{1}CTO Division, LG Electronics, Seoul, South Korea\\
\textsuperscript{2}David R. Cheriton School of Computer Science,
University of Waterloo, Waterloo, Canada\\
\textsuperscript{3}Department of Electrical Engineering,
KU Leuven, Leuven, Belgium\\
\textsuperscript{4}Computer Science and Engineering,
Ohio State University, Columbus, USA\\
\textsuperscript{5}Graduate School of Information Science and Technology,
The University of Osaka, Osaka, Japan
\\
}
}%
}

\vskip 0.3in

\input{mymacro}

\title{SilentWood: Efficient Private Inference \\Over Gradient Boosting Decision Forests}
\maketitle
\fancyhead{}                       
\renewcommand{\headrulewidth}{0pt} 

\begin{abstract}
Gradient boosting decision forests, used by XGBoost or AdaBoost, offer higher accuracy and lower training times than decision trees on large datasets.
Private inference protocols for decision trees can preserve both input and tree privacy.
However, naively extending them to decision forests by replication leads to impractical running times.
In this paper, we propose an efficient private decision inference protocol using homomorphic encryption.
We present several optimizations that identify and remove (approximate) duplication between trees, significantly reducing communication and computation costs over the naive approach.
We present the private inference protocol for highly scalable gradient boosting decision forests.
Our protocol (\sysname) is faster than parallel \revise{RCC-PDTE by up to 42.5x, Zama's Concrete ML XGBoost by up to 27.8x, and SoK-GGG's two-party garbled circuit protocol by 2.94x.}
\end{abstract}

\section{Introduction}
\label{sec:introduction}
\input{10-introduction}

\section{Preliminaries}
\label{sec:background}
\input{20-background}

\section{Methodology}
\label{sec:design}
\input{40-design}

\section{Evaluation}
\label{sec:evaluation}
\input{60-evaluation}

\section{Related Work}
\label{sec:related}
\input{80-related}

\section{Conclusion}
\label{sec:conclusion}
\input{90-conclusion}



\section*{AI Use Statement}
In this work, we used generative AI tools to formulate mathematical
claims, assist in writing proofs, implement methods, and review the
manuscript for numerical, notational, and textual inconsistencies.
We did not use generative AI tools to develop theoretical models or
conceptual frameworks; provide critical ingredients for proving
mathematical claims; generate synthetic datasets; propose or refine
hypotheses; design or provide feedback on research methodology or
experiments; assist with translation; clean or reformat datasets;
support qualitative or thematic data analysis; or interpret results.
We reviewed all AI-assisted work and take responsibility for the final content.

\section*{Ethics Statement}

Our work studies privacy-preserving inference for gradient boosting
decision forests using only public UCI datasets~\cite{uci}, which contain
no personally identifying information. No attacks were conducted against
real systems, and no harm resulted from this study. As with any
cryptographic technique, misuse is possible: the confidentiality that
protects legitimate models could also conceal discriminatory or harmful
ones, which we discourage.

\section*{Reproducibility Statement}
 We will release the artifact to reproduce the experimental results illustrated in \autoref{sec:evaluation}.

\bibliography{bibfile}
\bibliographystyle{iclr2027_conference}

\newpage
\appendix
\input{100-appendix}

\end{document}

%% file: mymacro.tex

\newcommand{\fixme}[1]{\textcolor{red}{\textbf{FIXME!!!} #1}}

\newcommand{\ignore}[1]{}
\newcommand{\authnote}[2]{{\bf[#1: #2]}}

\newcommand{\RK}[1]{{\authnote{\textcolor{orange}{RK}}{\textcolor{blue}{#1}}}}
\newcommand{\FK}[1]{{\authnote{\textcolor{orange}{FK}}{\textcolor{brown}{#1}}}}
\newcommand{\RM}[1]{{\authnote{\textcolor{orange}{RM}}{\textcolor{orange}{#1}}}}

\newcommand{\IGNORE}[1]{{#1}}

\renewcommand{\RK}[1]{}
\renewcommand{\FK}[1]{}
\renewcommand{\RM}[1]{}
\renewcommand{\IGNORE}[1]{{}}

\newcommand{\revise}[1]{{{\textcolor{black}{#1}}}}

\newcommand{\revisee}[1]{{{\textcolor{white}{#1}}}}

\newcommand{\mycomment}[1]{}


\newcommand{\etal}{\textit{et al.}\xspace}
\newcommand{\naive}{na\"{\i}ve\xspace}
\newcommand{\Naive}{Na\"{\i}ve\xspace}
\newcommand{\naively}{na\"{\i}vely\xspace}
\newcommand{\Naively}{Na\"{\i}vely\xspace}
\newcommand{\topk}{top-$k$\xspace}
\newcommand{\Topk}{Top-$k$\xspace}

\newcommand{\ultraverse}{\textsf{Ultraverse}\xspace}

\newcommand{\sysname}{\textsf{\small SilentWood}\xspace}
\newcommand{\sysnametitle}{\textsf{SilentWood}\xspace}

\newcommand{\set}[1]{\{#1\}}

\newcommand{\pie}[1]{%
\begin{tikzpicture}
 \draw (0,0) circle (1ex);\fill (1ex,0) arc (0:#1:1ex) -- (0,0) -- cycle;
\end{tikzpicture}%
}

\newcommand{\para}[1]{\vspace{0.05in}\noindent{\bf{#1}}}

\definecolor{red-brown}{rgb}{0.75, 0, 0}

%% file: 10-introduction.tex
Machine Learning as a Service (MLaaS) allows large user groups to use machine learning models.
A service provider trains a model and offers users inference using this model. \revise{Some of such examples include recommendation systems, spam filtering, and object detection used in e-commerce services, social networking services, and smart home.} An advantage of such MLaaS is that the provider can control access to the model and easily update it when new training data is available.
In contrast, its disadvantage is the privacy concern of the users revealing their test data to the provider which may be sensitive or legally restricted.
Cryptography offers a compelling solution to this problem. For example, the provider can use fully homomorphic encryption \cite{gentry09,fv12} to \textit{blindly} process the user's input features in an encrypted form through the trained model and yield an encrypted prediction result, which only the user can decrypt. 
However, this solution usually has high computational costs.
Alternatively, the user and the provider can engage in a two-party computation \cite{yao82,yao86,goldreich09} \revise{where the user's input is computed through the model in a \textit{blinded} manner, during which the server's trained model parameters and the user's private data are not leaked to each other. However, a two-party computation usually incurs high communication costs.}

In this paper, we develop a new protocol for private inference over gradient boosting decision forests\revise{, which} are collections of decision trees whose results are combined into the final model prediction.
Gradient boosting forests offer a better trade-off between higher accuracy and lower training time than training a single decision tree for large datasets.
Moreover, they retain a critical advantage of decision trees over many other machine learning techniques; their predictions are explainable, helping to fulfill legal, social, and ethical requirements.

Many private inference protocols over decision trees exist
(\autoref{sec:related}).
However, extending these to gradient boosting decision forests comes with two challenges.
First, the results of the decision trees need to be combined into a single prediction result, requiring further computation.
This can particularly pose a problem for protocols using homomorphic encryption,
since these computations increase \revise{the multiplication overhead}, significantly increasing the \revise{runtime} of the protocols.
Second, even naively running several independent decision tree protocols without combining their results can exhaust the server's computing resources.
Therefore, optimizations are necessary to reduce the cost of executing multiple decision tree evaluations on the same input.

To solve these problems, we propose \sysname, a private inference protocol for gradient boosting that uses three novel techniques. First, to reduce the high computation overhead caused by frequent homomorphic computations in FHE-based PDTE (private decision tree evaluation), \sysname applies node clustering (i.e., grouping and weighted-averaging tree nodes having similar thresholds) to eliminate redundant computations (\autoref{subsec:cluster}).  Second, \sysname proposes the Blind Code Conversion (BCC) protocol to combine SumPath's depth-independent multiplicative cost with MultiplyPath's arithmetic compatibility for score aggregation (\autoref{subsec:bcc}). BCC is a lightweight two-party protocol without garbled circuits or secret sharing: the server pads and shuffles its intermediate ciphertext so its plaintext looks uniformly random to the client. The client's role is to decrypt it and convert the underlying plaintext values \textit{blindly} to help the server's subsequent computation of score aggregation. This process ensures the information in the intermediate ciphertext to be inaccessible from both the client and server, while realizing arithmetic compatibility of the server's score aggregation. Lastly, to reduce communication overhead caused by repetitive data encoding in FHE-based PDTE, \sysname's ciphertext compression carefully removes repetitive encodings before encryption and then homomorphically restores the original data format on the server side (i.e., homomorphic decompression), thus minimizing ciphertext size while preserving necessary SIMD comparisons (\autoref{subsec:compression}).

While RCC-PDTE~\cite{rcc-pdte} (\autoref{appendix:cw-comparison}), the state-of-the-art PDTE protocol, is effective for computing a single decision tree, naively repeating this protocol for all decision trees (100 in our default setup) can take more than 5 minutes for a single inference, which is prohibitively long.
The inference time of \sysname is, on average, faster than this baseline by 42.5x, faster than Zama's Concrete ML~\cite{concrete-ml} XGBoost by 34.0x, and faster than SoK-GGG (i.e., a state-of-the-art MPC protocol based on GC) by 2.94x (\autoref{sec:evaluation}).
Our paper's contributions are as follows:

\begin{itemize}

\item We propose three optimization techniques for efficient private inference of random forest models: node clustering, blind code conversion, and ciphertext compression.

\item We implemented \sysname, a prototype of our proposed techniques.

\item We provide evaluation results of \sysname and compare it to other state-of-the-art protocols.

\end{itemize}

%% file: 20-background.tex
\subsection{Gradient Boosting Models}
\label{subsec:boosted-decision-trees}

Ensemble decision trees~\cite{PRML_Bishop} combine many shallow trees (weak learners) to improve prediction, typically via either random forests~\cite{random-forest} or gradient boosting~\cite{gradient-boosting}. Random forests train trees independently and aggregate their outputs (e.g., voting/averaging), whereas gradient boosting trains trees sequentially to correct prior residual errors, yielding a strong additive model. We focus on private inference for AdaBoost~\cite{AdaBoost} and XGBoost~\cite{XGBoost}. In AdaBoost, each tree has a weight and each leaf outputs a signed value ((+)/(-)); inference traverses every tree by threshold comparisons, retrieves the reached leaf value, multiplies it by the tree weight, sums across trees, and predicts by the sign of the final sum. XGBoost similarly sums leaf values of traversed trees: for binary classification it applies a sigmoid to the sum and thresholds at 0.5; for multi-class it computes a per-class sum, applies softmax, and outputs the argmax.

\subsection{Fully Homomorphic Encryption (FHE)} 
\label{subsec:fhe}

Fully Homomorphic Encryption (FHE)~\cite{fhe} enables addition and multiplication directly on ciphertexts, allowing computation on encrypted data without decryption. \sysname adopts an RLWE-family scheme, BFV~\cite{fan2012somewhat}, primarily for efficient SIMD batching: an $n$-dimensional plaintext vector is encoded (\textsf{Encode}) into a $(n-1)$-degree (or lesser degree) polynomial $M(X)$, then encrypted (\textsf{Encrypt}) under a secret polynomial $S(X)$ to an RLWE ciphertext (with a large scaling factor $\Delta$ as part of the scheme). The scheme supports cipher-plaintext and cipher-cipher addition/multiplication (\textsf{PAdd/CAdd/PMult/CMult}) as well as slot rotations (\textsf{Rotate}) that cyclically permute packed plaintext elements without decryption. SIMD means these operations act element-wise over all packed slots at once; typical slot counts are $2^{13} \sim 2^{15}$, providing substantial parallelism. We will treat these FHE primitives as black boxes to implement private gradient-boosting inference. The supported operations of the BFV scheme are detailed in \autoref{appendix:bfv-scheme}.

\subsection{Private Decision Tree Evaluation}
\label{subsec:pdte}

Private Decision Tree Evaluation (PDTE) is a protocol between a server holding a decision tree and a client holding a feature vector for one sample. At the end of PDTE, the server learns nothing about the client's inputs, while the client learns nothing about the server's model beyond the prediction result and model hyper-parameters (e.g., feature/class types)\footnote{\revise{More precisely, PDTE's server-side privacy goal is to reveal no more model information than plaintext inference.}}. The server typically returns class scores (probabilities), from which the client selects the top class (or applies a tie-breaking rule). PDTE generally consists of two steps: (i) value comparison at each node; and (ii) tree path evaluation.

\para{Value Comparison at Each Node:} Tree traversal requires \textit{greater-than} comparisons between an input feature $X$ and a node's threshold $Y$. Under FHE, only two arithmetic operations $(+, \cdot)$ are natively supported, so comparisons must be implemented as an arithmetic circuit over $(+, \cdot)$. A direct approach is to bitwise-decompose $X, Y$ into encrypted bits and \textit{homomorphically} evaluate a standard comparator circuit, where all logic gates are implemented as bit-wise $(+, \cdot)$ operations of encrypted bits. This approach computationally requires a cipher-cipher multiplication (\textsf{CMult}) chain as long as the bit length of $X, Y$. In FHE, cipher-cipher multiplication is computationally heavy, so shortening the multiplication chain is crucial. To this end, a more efficient method of value comparison is based on constant-weight (CW) encoding~\cite{rcc-pdte}, where numbers are encoded using a fixed Hamming weight, and the \textsf{CMult} chain's length for value comparison is equivalent to the Hamming weight. While \sysname is compatible with both methods, our prototype adopts the CW-based one. We explain the details of these two comparison methods in \autoref{appendix:cw-comparison}.

\para{Tree Path Evaluation:} Given encrypted comparison results at all nodes (1 if $>$ is satisfied; 0 otherwise), PDTE's next step is to determine the reached leaf by evaluating path conditions. Two common methods are MultiplyPath (path conjugation)~\cite{knlas19} and SumPath~\cite{tkk19}. As illustrated in \autoref{fig:path-formula} (\autoref{appendix:edge-formulas}), MultiplyPath assigns each edge a 0/1 indicator and multiplies indicators along a root-to-leaf path, so only the true path yields 1 (others 0). However, this requires a cipher-cipher multiplication ($\textsf{CMult}$) chain length proportional to tree depth. On the other hand, SumPath assigns 0 to satisfied edges and 1 to unsatisfied edges and sums along each path, so the true path yields 0 and every other path yields its number of unsatisfied edges ($1\!\sim\!d$); each sum is then multiplied by a fresh random non-zero value to hide the count. This needs only ciphertext additions and one plaintext multiplication, which is substantially cheaper under FHE. After homomorphically computing the values for all tree paths, the server returns them to the client, who decrypts and selects the unique leaf whose SumPath equals 0 (or MultiplyPath equals 1) to obtain the prediction. We provide an example of tree path evaluation based on MultiplyPath and SumPath in ~\autoref{fig:path-computation} (\autoref{appendix:multiplypath-sumpath}).

%% file: 40-design.tex
This section explains the problems of applying existing PDTE methods to evaluate private gradient boosting models, and how \sysname solves them through three novel techniques: node clustering, blind code conversion, and ciphertext compression.

\para{Threat Model:} Our threat model for privacy assumes a semi-honest client (i.e., follows the protocol) and a malicious server (i.e., may deviate from the protocol)\footnote{For assuming a malicious client and a malicious server, see \autoref{appendix:blind-shuffling-analysis}}. They try to learn about each other's secrets from exchanged messages. The server's secret is its trained gradient boosting model; the client's secret is its input feature values. Their common knowledge is the feature names and feature bit-length. At the end of the protocol, the server learns nothing about the client's features; the client learns only the inference scores/probabilities for each class and the fact that the number of tree paths is capped at some multiple of the cyclotomic polynomial's degree (e.g., $2^{14}$) as a hyperparameter. Besides that, the client learns nothing about the server's model (e.g., node/tree count, depth, or node thresholds).

\subsection{Node Clustering}
\label{subsec:cluster}

\begin{figure}[t]
\centering

\begin{minipage}[c]{0.47\linewidth}
\centering

\includegraphics[width=0.90\linewidth]{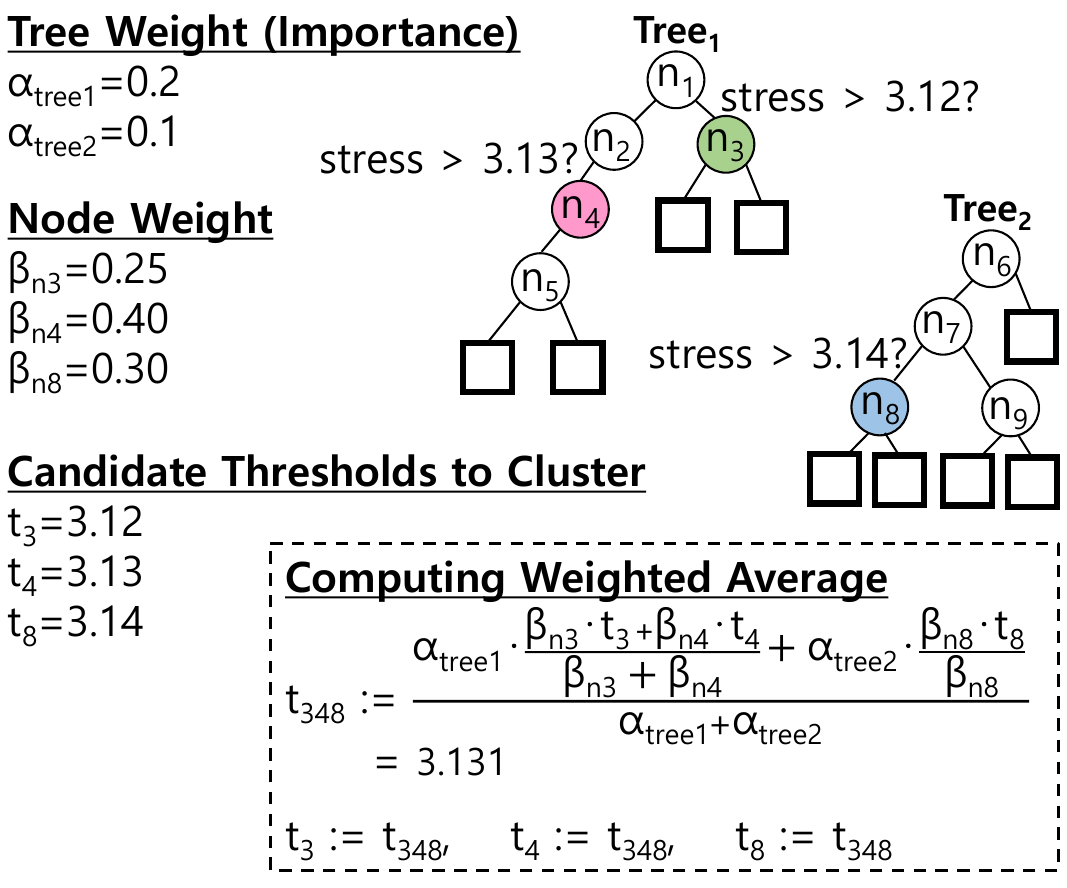}

\captionof{figure}{An example of node clustering}
\label{fig:node-clustering}

\end{minipage}
\hfill
\begin{minipage}[c]{0.51\linewidth}
\scriptsize

\hrule
\vspace{2pt}

\refstepcounter{algorithm}
\label{alg:node-clustering}
{\textbf{Algorithm \thealgorithm}\quad Accuracy-preserving merge\par}

\vspace{2pt}
\hrule
\vspace{3pt}

\begin{algorithmic}[1]
\STATE \textbf{Input:} $\mathcal{N}, s$
\textcolor{red-brown}{$\rhd$ nodes, clustering intensity}

\REPEAT
  \STATE $\mathit{changed} \gets \textbf{false}$
  \FOR{$n_i \in \mathcal{N}$}

\STATE $\mathcal{C} \gets \left\{\,
\begin{array}{@{}l@{}}
n \in \mathcal{N} \mid
n.\mathit{name}=n_i.\mathit{name}\ \wedge \\[1pt]
\qquad
\dfrac{|n.\mathit{threshold}-n_i.\mathit{threshold}|}
{\textbf{MAX}(n_i.\mathit{name})-
 \textbf{MIN}(n_i.\mathit{name})} \leq s
\end{array}
\,\right\}$
    \IF{$| \{ n.\mathit{threshold} \mid n \in \mathcal C \} | > 1$}
      \STATE $\mathit{new\_threshold}
    \gets \textsf{WeightedAverage}(\mathcal C)$
      \FOR{$n \in \mathcal{C}$}
        \STATE $n.\mathit{threshold}
        \gets \mathit{new\_threshold}$
      \ENDFOR
      \STATE $\mathit{changed} \gets \textbf{true}$
    \ENDIF
  \ENDFOR

\UNTIL{$\neg\mathit{changed}$}
\textcolor{red-brown}{$\rhd$ stop when no more clustering occurs}

\STATE \textbf{Output:} $\mathcal{N}$
\textcolor{red-brown}{$\rhd$ clustered final tree nodes}
\end{algorithmic}

\vspace{2pt}
\hrule

\end{minipage}

\end{figure}

\textit{Problem of Many Tree Nodes and Paths.}
Existing PDTE protocols homomorphically compare each internal node's threshold $Y$ with the client’s encrypted feature value $X$ by homomorphically evaluating the value comparison logic for $X > Y$ (see \autoref{fig:cw-algorithm}), which entails expensive \textsf{CMult}, \textsf{PMult}, and \textsf{Rotate} operations.
This cost becomes a substantial bottleneck for private inference over gradient-boosted forests, since such models typically comprise large ensembles
(e.g., 100$\sim$1000 trees), involving orders of magnitude more nodes and paths than a single decision tree.

\sysname's insight is to homomorphically compute the comparison logic once per distinct $\langle$\textit{feature},\textit{threshold}$\rangle$ value pair across all trees and reuse them for each group of nodes having the same value pair. This helps cut down the computation overhead for the nodes whose thresholds are exactly identical. To further reduce the overhead, \sysname clusters nodes that share the same feature name and \emph{similar} thresholds, then replaces the thresholds in the cluster with their \textit{weighted} average: weighted by the tree weight $\alpha$ and the node weight $\beta$. In \autoref{fig:node-clustering}'s example, nodes $n_3,n_4,n_8$ (e.g., stress level) have close thresholds: 3.14, 3.12, and 3.13. \sysname clusters these nodes together and updates their thresholds to 3.131 as their weighted average (dashed-box formula). Now their comparisons are identical, so we compute the comparison for these 3 nodes once instead of three times. Meanwhile, since changing thresholds can alter accuracy, \sysname addresses this by using: (i) clustering intensity adjustment, (ii) weighted averaging, (iii) leaf value re-optimization; and (iv) accuracy-preserving merge.

\para{Clustering Intensity:} With learnable intensity $s$, we co-cluster node $n_i$ and $n_j$ if they share the same feature name and their values are close enough such that $\frac{|\mathit{n_i.threshold} - \mathit{n_j.threshold}|}{\textbf{MAX}(\mathit{n_i.name}) - \textbf{MIN}(\mathit{n_i.name})} \leq s$, where $\textbf{MAX}(\mathit{n_i.name})$ and $\textbf{MIN}(\mathit{n_i.name})$ are the feature $i$'s max/min value over the whole training data. The optimal $s$ is dataset-dependent.

\para{Weighted Threshold Average:} For a given node cluster $\mathcal{C}$, we update the node thresholds in it to a weighted average by using tree weight $\alpha$ and node weight $\beta$. The high-level idea is that more influential nodes/trees should change their thresholds less to preserve accuracy. To compute tree $i$'s weight $\alpha_i$, we first compute $w_i$, the sum of hit absolute leaf values over all training samples on tree $i$, and normalize it over all $T$ trees in the model as $\alpha_i = \frac{w_i}{\sum_{k=1}^T w_k}$. We compute node $j$'s weight $\beta_j$ as the fraction of information gain of node $j$ among all nodes in $\mathcal{C}$ that belong to this tree. Using all $\alpha_i$ and $\beta_j$, we scale thresholds and get the weighted average as Equation~\eqref{eq:threshold-avg}.
\begin{equation}
\label{eq:threshold-avg}
t_{\mathit{average}} = \dfrac{\sum_{i=1}^{T} \alpha_i \cdot \frac{ \sum \{ \beta_j \cdot t_j \, | \, n_j \in \mathit{Tree}_i \, \wedge \, n_j \in \mathcal{C} \} }{\sum \{ \beta_j \, | \, n_j \in \mathit{Tree}_i \, \wedge \, n_j \in \mathcal{C} \} } }{\sum_{i=1}^{T} \alpha_i}
\end{equation}

\para{Leaf Value Re-optimization:}
After updating thresholds for a cluster $\mathcal{C}$, we re-optimize the leaf values (scores) while keeping the tree topology fixed.
Concretely, we re-route the training samples under the updated thresholds and, for each leaf $\ell$, aggregate the first-order and second-order loss derivatives with respect to the raw prediction $\hat{y}(x)$: $G_\ell = \sum_{x \in \ell} \frac{\partial}{\partial \hat{y}(x)} \mathcal{L}\!\big(y(x), \hat{y}(x)\big)$,
$H_\ell = \sum_{x \in \ell} \frac{\partial^2}{\partial \hat{y}(x)^2} \mathcal{L}\!\big(y(x), \hat{y}(x)\big)$.
We then apply a regularized Newton update to the leaf value:
$v_\ell \leftarrow v_\ell - \eta \cdot \frac{G_\ell}{H_\ell + \lambda}$, where $\lambda$ is the $\ell_2$ regularization coefficient on leaf values and $\eta$ is the learning rate.
This update recalibrates the existing leaf outputs under the modified decision boundaries induced by threshold averaging.

\para{Accuracy-preserving Merge:} Once the candidate cluster $\mathcal{C}$'s thresholds are weight-averaged and leaves are re-calibrated as described in Algorithm~\autoref{alg:node-clustering}, we decide whether to commit/abort it. Basically, we evaluate the updated model on the \revise{validation} set: if validation accuracy is unchanged or improved, we commit; otherwise we abort. We repeat this in a bottom-up manner until no more merge-able thresholds. Our evaluation (\autoref{tab:result-node-clustering-accuracy}) shows that a high $s$ does not reduce validation accuracy because accuracy-degrading merges are rejected.

\para{Applying Node Clustering to MPC:} Fewer distinct comparisons reduce cost in encrypted inference beyond FHE. In MPC, comparisons are often dominant; e.g., \autoref{tab:sok-breakdown} (\autoref{appendix:more-exp}) shows GC-based XGBoost spends 84.1\% time on comparisons. Thus, node clustering can directly lower MPC runtime and is compatible with both FHE and MPC.

\subsection{Blind Code Conversion (BCC)}
\label{subsec:bcc}

\begin{table}[t]
\setlength{\tabcolsep}{2pt}
\footnotesize
\centering

\begin{minipage}[t]{0.49\linewidth}
\vspace{0pt}
\centering

\resizebox{\linewidth}{!}{%
\begin{tabular}{|l|}
\hline
\textbf{\underline{Notations}}\\

\begin{tabular}{p{0.7cm} p{7.5cm}}
$N$ &: The number of plaintext slots in an RLWE ciphertext\\
$\mathbb{Z}_p, \mathbb{Z}_p^{*}$ &: The plaintext space $\{0, \cdots, p-1\}$ ($p$ prime); $\mathbb{Z}_p$ without 0\\
$\mathcal{U}_p$ &: A custom conversion table which defines the mapping of \\
& \textcolor{white}{ : } each plain value $\{0, \cdots, p-1\}$ to some $\{u_0, u_1, \cdots, u_{p-1}\}$\\
$f$&: The public zero/non-zero distribution of $\vec{w}$\\
$f_{\mathit{zero}}$ &: The number of 0s in each SumPath list as hyperparameter\\
$\vec{v}^{(i)}$ &: An array whose slot index 0 stores the $i$-th path cost $c_i$\\
$\vec{\rho}$ &: A mask array uniformly drawn from $\mathbb{Z}_p^{*}$\\
$\vec{u}$ &: An intermediate array that stores shuffled SumPath values\\
$\vec{w}$ &: A shuffled array of $f_{\mathit{zero}}$ 0s and $(N-f_{\mathit{zero}})$ uniform non-zero values \\
$\vec{c}$ &: The client converts $\vec{w}$ into $\vec{c}$ according to table $\mathcal{U}_p$\\
$\llangle \vec{v}^{(i)} \rrangle$ &: A ciphertext encrypting $\vec{v}^{(i)}$
\textcolor{red-brown}{$\rhd$ the same applies to: $\vec{u}, \vec{w}, \vec{c}$}\\
\end{tabular}

\\\hline
\end{tabular}%
}

\end{minipage}
\hfill
\begin{minipage}[t]{0.49\linewidth}
\vspace{0pt}
\centering

\resizebox{\linewidth}{!}{%
\begin{tabular}{|l|}
\hline
\textbf{\underline{Protocol}}\\

\begin{tabular}{p{0.8cm} p{7.4cm}}
\textbf{{Input}} &: $\bm\{\llangle \vec{v}^{(i)} \rrangle\bm\}_{i=1}^\ell$
\textcolor{red-brown}{$\rhd$ a list of $\ell$ ciphertexts storing all $\ell$ SumPaths}\\

\textbf{{Server}} &: 1) Randomly rotate each
$\bm\{\llangle \vec{v}^{(i)} \rrangle\bm\}_{i=1}^\ell$
and merge as $\llangle \vec{u} \rrangle$.\\

&\textcolor{white}{:} 2) $\llangle \vec{u} \rrangle \gets \textsf{PMult}(\llangle \vec{u} \rrangle, \textsf{Encode}(\vec{\rho}))$, \; $\vec{\rho} \xleftarrow{\$} (\mathbb{Z}_p^{*})^N$.\\

&\textcolor{white}{:} 3) Create a dummy array $\vec{d}$ (non-zero entries $\xleftarrow{\$} \mathbb{Z}_p^{*}$) such\\

&\textcolor{white}{: 3)}  that  the array $\vec{u} + \vec{d}$ has exactly $f_{\mathit{zero}}$ zeros\\

&\textcolor{white}{:} 4)
$\llangle \vec{w} \rrangle
\gets
\textsf{PAdd}(\llangle \vec{u} \rrangle,\textsf{Encode}(\vec{d}))$.\\

\textbf{S}$\rightarrow$\textbf{C} &: Send $\llangle \vec{w} \rrangle$ and $\mathcal{U}_p$.\\

\textbf{Client} &: 1) Decrypt $\llangle \vec{w} \rrangle$ to $\vec{w}$.\\

& \textcolor{white}{:} 2) Create $\vec{c}$ by converting each value in $\vec{w}$ according to $\mathcal{U}_p$.\\

& \textcolor{white}{:} 3) Encrypt $\vec{c}$ to $\llangle \vec{c} \rrangle$.\\

\textbf{C}$\rightarrow$\textbf{S} &: Send $\llangle \vec{c} \rrangle$.\\

\textbf{{Output}} &: $\llangle \vec{c} \rrangle$\\
\end{tabular}

\\\hline
\end{tabular}%
}

\end{minipage}

\caption{The blind code conversion (BCC) protocol}
\label{tab:bcc}
\end{table}

\textit{Problem of Final Score Aggregation.} In a gradient boosting forest, the final prediction score is the sum of leaf values on the \emph{true} paths of all trees. When it comes to private evaluation of this, the computation should be compatible with FHE arithmetic ($+,\cdot$) and also efficient. Unfortunately, existing PDTE methods (\autoref{subsec:pdte}) do not meet both. With SumPath~\cite{rcc-pdte}, a leaf's SumPath is $0$ on the true path and a masked non-zero value $\rho_j c_j$ otherwise (\autoref{subsec:pdte}), but there is no low-depth arithmetic expression over $(+,\cdot)$ to express the logic: $\textsf{SumPath==0 ? leafValue : 0}$. On the other hand, MultiplyPath~\cite{cdpp22} can compatibly express this logic as $\textsf{MultiplyPath}\cdot\textsf{LeafValue}$, but it requires a cipher-cipher multiplication chain (\textsf{CMult}) as long as the maximum depth of all tree paths, which is computationally expensive.

\sysname's novelty is resolving the dilemma of SumPath and MultiplyPath by introducing a light-weight two-party blind code conversion (BCC) protocol (consuming 1 RTT), as formally described in \autoref{tab:bcc}. This protocol essentially lets the client blindly convert SumPath values into MultiplyPath values for arithmetic compatibility. Initially (after path evaluation), the server packs the encrypted path costs into a list of intermediate ciphertexts ($\bm\{\llangle \vec{v}^{(i)} \rrangle\bm\}_{i=1}^\ell$), homomorphically (blindly) shuffles their slots, merges them as a single ciphertext $\llangle \vec{u} \rrangle$, applies the per-leaf masks $\vec{\rho}$ with a single \textsf{PMult}, and homomorphically pads it with dummy values to produce the final uniformly shuffled ciphertext $\llangle \vec{w} \rrangle$. At this point, the plaintext of $\llangle \vec{w} \rrangle$ has distribution $f$: a uniform mixture of zero : non-zero ratio of $1\!:\!1$. Since $p$ is prime and $1 \le c_j \le d < p$, every false-path value $\rho_j c_j$ is non-zero and uniform over $\mathbb{Z}_p^{*}$, so exactly the $T$ true paths are zero (\autoref{appendix:blind-shuffling-analysis}). Next, the server sends $\llangle \vec{w} \rrangle$ with a conversion table $\mathcal{U}_p$ to the client, who in turn decrypts it, converts $0\mapsto 1$ and non-zero $\mapsto 0$, re-encrypts them as $\llangle \vec{c} \rrangle$, and returns it to the server. Finally, the server multiplies (\textsf{PMult}) the received ciphertext $\llangle \vec{c} \rrangle$ with a plaintext leaf-score array, whose slot indices are aligned with the shuffled ones in $\llangle \vec{c} \rrangle$, and aggregates per-class prediction score by homomorphic rotations/additions (\textsf{Rotate}, \textsf{CAdd}). 

We provide an example of the BCC protocol in \autoref{appendix:bcc-example}. BCC contributes to the inference speedup by 19.3x (\autoref{tab:result-generic}, \autoref{tab:speedup-factors})

\begin{figure}[t]
\centering

\begin{sidealgorithm}{Blind Shuffling}{alg:blind-shuffling}
\begin{algorithmic}[1]
\STATE {\textbf{Input:} $T, \, \bm\{ \llangle \vec{v}^{(i)} \rrangle\bm\}_{i=1}^{\ell}, \,  \bm\{ s_i \bm\}_{i=1}^{\ell}$ }
\STATE { \textcolor{red-brown}{$\rhd$ total number of trees, encrypted path costs, plain leaf scores}}
\STATE {$\vec{\sigma} \gets \textit{genDist(f)}$ \textcolor{red-brown}{$\rhd$ An offset list for random $N$-slot permutation}}
\STATE {$\vec{x} \gets 0 $, \; $\vec{d} \gets 0$ \; \textcolor{red-brown}{$\rhd$ leaf-score array}} \; \textcolor{red-brown}{$\rhd$ dummy-SumPath array}
\STATE {$
\llangle \vec{u} \rrangle \gets \llangle [0, 0, \ldots, 0] \rrangle$ \textcolor{red-brown}{$\rhd$ a ciphertext of shuffled SumPaths}}
\FOR{$i \text{ } \textbf{in} \text{ } \{1, 2, \ldots, \ell \}$  \textcolor{red-brown}{$\rhd$ permute each $i$-th SumPath}}
    \STATE {$\llangle \vec{u} \rrangle \gets \textsf{CAdd}\bm(\llangle \vec{u} \rrangle, \textsf{Rotate}\bm( \llangle\vec{v}^{(i)} \rrangle, \sigma_i  \bm) \bm)$}
    \STATE {$x_{\sigma_i} \gets s_i$ } \textcolor{red-brown}{$\rhd$ also, store the slot-aligned leaf score}
\ENDFOR
\STATE {$\vec{\rho} \xleftarrow{\$} (\mathbb{Z}_p^{*})^{N}$}
\STATE {$\llangle \vec{u} \rrangle \gets \textsf{PMult}\bm(\llangle \vec{u} \rrangle, \textsf{Encode}(\vec{\rho})\bm)$} \textcolor{red-brown}{$\rhd$ $\rho_j c_j \neq 0$ iff false path}
\FOR{$i \text{ } \textbf{in} \text{ } \{\ell + 1, \ldots, N - (f_{\textit{zero}} - T) \}$ \textcolor{red-brown}{$\rhd$ $f_{\textit{zero}}$: freq. of 0s}}
    \STATE {$d_{\sigma_i} \xleftarrow{\$} \mathbb{Z}_p^{*}$ \textcolor{red-brown}{$\rhd$ permute dummy non-zero SumPaths}}
\ENDFOR
\STATE {$\llangle \vec{w} \rrangle \gets \textsf{PAdd}\bm(\llangle \vec{u} \rrangle, \textsf{Encode}(\vec{d}) \bm)$ \textcolor{red-brown}{$\rhd$ add dummy SumPaths}}
\STATE {\textbf{Output:} $\llangle \vec{w} \rrangle$, $\vec{x}$ \textcolor{red-brown}{$\rhd$ shuffled SumPaths, slot-aligned leaf scores}}
\end{algorithmic}
\end{sidealgorithm}
\hfill
\begin{sidealgorithm}{Blind Score Aggregation}{alg:blind-scoring}
\begin{algorithmic}[1]
\STATE {\textbf{Input:} $\llangle \vec{c} \rrangle, \vec{x}$ \textcolor{red-brown}{$\rhd$ client-converted SumPaths, aligned leaf scores}}
\STATE {$\llangle \vec{z} \rrangle \gets \textsf{PMult}\bm( \llangle \vec{c} \rrangle, \textsf{Encode}\bm(\vec{x}\bm) \bm)$ \textcolor{red-brown}{$\rhd$ zero out false-path scores}}
\FOR{$n \text{ } \textbf{in} \text{ } \{\log_2 N, (\log_2 N)-1, \, \ldots, \, 2, 1 \}$}
    \STATE {$\vec{m}^{(1)} \gets [\overbrace{1, \ldots, 1}^{2^{n-1} \text{ slots}}, \overbrace{0, \ldots, 0}^{2^{n-1} \text{slots}}, \overbrace{0, \ldots, 0}^{N - 2^n \text{ slots}}] $ \textcolor{red-brown}{$\rhd$ 1st $2^{n-1}$ slot filter}}
    \STATE {$\vec{m}^{(2)} \gets [\overbrace{0, \ldots, 0}^{2^{n-1} \text{ slots}}, \overbrace{1, \ldots, 1}^{2^{n-1} \text{slots}}, \overbrace{0, \ldots, 0}^{N - 2^n \text{ slots}}] $  \textcolor{red-brown}{$\rhd$ 2nd $2^{n-1}$ slot filter}}
    \STATE {$\llangle \vec{z}^{(1)} \rrangle = \textsf{PMult}\bm( \llangle \vec{z} \rrangle, \textsf{Encode}\bm(\vec{m}^{(1)}\bm) \bm) $}
    \STATE {$\llangle \vec{z}^{(2)} \rrangle = \textsf{PMult}\bm( \llangle \vec{z} \rrangle, \textsf{Encode}\bm(\vec{m}^{(2)} \bm)\bm) $}
    \STATE {$\llangle \vec{z} \rrangle = \textsf{CAdd}\bm( \llangle \vec{z}^{(1)} \rrangle, \textsf{Rotate}\bm( \llangle \vec{z}^{(2)} \rrangle, 2^{n-1} \bm) \bm)$}  \textcolor{red-brown}{$\rhd$ align, add}
\ENDFOR

\STATE {\textbf{Output:} $\llangle \vec{z} \rrangle$  \textcolor{red-brown}{$\rhd$ slot 0 stores the aggregated prediction score}}
\end{algorithmic}
\end{sidealgorithm}

\end{figure}

\para{Blind Shuffling:} Algorithm~\autoref{alg:blind-shuffling} describes how the server homomorphically pads and shuffles SumPaths according to a random permutation plan $\vec{\sigma}$ whose value distribution is $f$ (line 3). Following the random $\vec{\sigma}$, the server accordingly rotates each real path cost (line 6$\sim$9), masks the merged ciphertext (line 10), adds randomly placed dummy SumPaths (line 11$\sim$14), and compactly merges all of them in the final ciphertext $\llangle \vec{w} \rrangle$. Once the client receives $\llangle \vec{w} \rrangle$ and returns its converted (i.e, $0\mapsto 1$, non-zero $\mapsto 0$) version $\llangle \vec{c} \rrangle$, the server homomorphically filters and aggregates the valid leaf scores in it as described in Algorithm~\autoref{alg:blind-scoring}: the server multiplies $\llangle \vec{c} \rrangle$ with the slot-aligned plaintext leaf score array $\vec{x}$ to homomorphically zero out the leaf scores for false paths as well as dummy SumPaths (line 2), then sums the filtered scores and places them at slot 0 of $\llangle \vec{z} \rrangle$ (line 3$\sim$9).


BCC is computationally efficient because it requires no cipher-cipher multiplication (\textsf{CMult}) chain. Unlike FHE+GC hybrids, it needs no garbled circuits or secret sharing: one RTT with a few packed FHE ciphertexts suffices even for hundreds of trees, since extra paths only fill spare slots. For security, the server's masking, padding, and uniform shuffling make the client's view of SumPaths to be distribution $f$ (Lemma~\ref{lem:view}), while the server never sees their decryption. Complexity and formal analyses are in \autoref{appendix:bcc-complexity} and \autoref{appendix:blind-shuffling-analysis}.

\para{Extension to AdaBoost and Others:} Private AdaBoost evaluation can be implemented by replacing each leaf score by $+\omega$ or $-\omega$ (class sign) where $\omega$ is the tree weight. Its computation time and complexity are unchanged. In general, BCC can be used for any FHE applications where the true frequency distribution of values for programmable bootstrapping in a target ciphertext is the server's private information to be hidden from the client (e.g., the number of zero SumPaths, since it equals/leaks the total number of trees in XGBoost).

\subsection{Ciphertext Compression}
\label{subsec:compression}

\begin{algorithm}[t]
\scriptsize
\begin{algorithmic}[1]
\STATE {\textbf{Input:} $\{A_0,\ldots,A_{M-1}\}$, $\{\mathit{start}_j\}$, $\{\mathit{gap}_0,\ldots,\mathit{gap}_{R-1}\}$, $R$ \\ \textcolor{red-brown}{$\rhd$ $M$ data arrays (one per encoded bit), feature $j$'s start slot, $r$-th repetition offset, repetitions per ciphertext}}
\FOR{$k \in \{0, \ldots, \lfloor M/R \rfloor - 1\}$ \textcolor{red-brown}{$\rhd$ every $R$ arrays share one ciphertext}}
    \STATE {$P \gets [0, \ldots, 0]$ \textcolor{red-brown}{$\rhd$ one compact plaintext array}}
    \FOR{$r \in \{0, \ldots, R-1\}$ \textcolor{red-brown}{$\rhd$ $r$-th array of the group}}
        \STATE {$P[\mathit{start}_j + \mathit{gap}_r] \gets A_{kR+r}[j]$ \textbf{for all} $j$ \textcolor{red-brown}{$\rhd$ the slot a repetition would have used now holds another array}}
    \ENDFOR
    \STATE {$C_k \gets \textsf{Encrypt}(\textsf{Encode}(P), \mathit{sk})$}
\ENDFOR
\STATE {\textbf{Output:} $\{C_0, \ldots, C_{\lfloor M/R \rfloor - 1}\}$ \textcolor{red-brown}{$\rhd$ $M/R$ ciphertexts instead of $M$}}
\end{algorithmic}
\caption{Ciphertext Compression}
\label{alg:ciphertext-compression}
\end{algorithm}
\textit{Problem of Huge Communication Cost.} In FHE-based PDTE, ciphertext size grows quickly with the number of nodes because inputs are encoded/encrypted bit-wise. For example, RCC-PDTE~\cite{rcc-pdte} requires an 85.7\,MB query ciphertext for 32-bit feature precision, where network transmission dominates inference time. The major bottleneck is \emph{repetitive data encoding} for SIMD comparisons, which intentionally duplicates the same feature value across multiple slots for the server's computational efficiency, but instead sacrifices slot utility.
For example, suppose that the server's tree model has three nodes on \textit{age} (thresholds 21, 23, 45) and two on \textit{sleep} (6, 8). To homomorphically compare against the encrypted client inputs \textit{age}$=28$ and \textit{sleep}$=7$ in SIMD, the client repeatedly encodes \textit{age} and \textit{sleep} three times each (capped by the maximum repetition of all features), wasting slots for redundancy.

To reduce such communication inefficiency, \sysname introduces \textit{ciphertext compression}: (1) the client removes such repetitions \emph{before} encryption to form compact ciphertexts; (2) upon receiving them, the server \emph{homomorphically} decompresses to restore the intended repetition pattern. Algorithm~\autoref{alg:ciphertext-compression} generalizes the client's packing based on: (1) the $M$ plaintext data arrays $A_0,\ldots,A_{M-1}$, one per CW-encoded bit position; (2) each feature $j$'s start slot $\mathit{start}_j$; (3) the offsets $\mathit{gap}_0,\ldots,\mathit{gap}_{R-1}$ at which the $R$ repetitions of a feature would have been encoded; and (4) the number $R$ of arrays packed into one ciphertext, so that the slot a repetition would have occupied now holds the bit of another array. The key requirement is that the offsets $\mathit{gap}_r$ are the same for every feature, so the server can restore all repetitions \revise{homogeneously} in SIMD (e.g., \textit{age} and \textit{sleep} are expanded simultaneously).

The server's homomorphic decompression is essentially a homomorphic reverse operation of Algorithm~\autoref{alg:ciphertext-compression}. Ciphertext (de-)compression does not leak the client's input data to the server. We provide an example of ciphertext compression and the ciphertext decompression algorithm in \autoref{appendix:ciphertext-decompression}. 


\begin{table*}[h]
\setlength{\tabcolsep}{2.5pt}
\scriptsize
\renewcommand\arraystretch{1.2}
\centering

\begin{tabular}{|l||rr||r|rr|rr||r|rr|rr|}
\toprule
\multirow{3}{*}{
{\textbf{Benchmark}}}
 & \multicolumn{2}{c||}{{\textbf {MPC}}} & \multicolumn{5}{c||}{{\textbf {CPU-FHE}}} & \multicolumn{5}{c|}{{\textbf {GPU-FHE}}} \\   
  \cmidrule{2-3} \cmidrule{4-8} \cmidrule{9-13} & \multicolumn{2}{c||}{{\textbf{SoK-GGG}}} &  \multicolumn{1}{c|}{{\textbf{Zama}}} & \multicolumn{2}{c|}{{\textbf{Baseline}}} & \multicolumn{2}{c||}{\textbf{\sysname}} & \multicolumn{1}{c|}{{\textbf{Zama}}} & \multicolumn{2}{c|}{{\textbf {Baseline}}} & \multicolumn{2}{c|}{\textbf{\sysname}} \\ 
\cmidrule{2-3} \cmidrule{4-4} \cmidrule{5-6} \cmidrule{7-8} \cmidrule{9-9} \cmidrule{10-11} \cmidrule{12-13} & \textsf{16-bit} & \textsf{32-bit} & \textsf{14-bit} & \textsf{16-bit} & \textsf{32-bit} & \textsf{16-bit} & \textsf{32-bit} & \textsf{14-bit} & \textsf{16-bit} & \textsf{32-bit} & \textsf{16-bit} & \textsf{32-bit} \\
\hline
Spam 		 & 5.98s &	5.95s & 72.68s& 153.68s  	& 162.53s 	& 3.82s  & 5.26s & 47.42s & N/A &  N/A &  1.65s &  2.27s\\
Steel 		& 5.41s &	5.56s & 18.01s & 37.24s  	& 38.32s   	& 1.81s  & 1.98s & 10.57s & 3.86s &  5.62s &  1.04s &  1.21s\\
Breast 		 & 5.24s &	5.69s & 12.32s & 20.62s   	& 28.27s  	& 1.57s  & 1.81s & 9.08s & 2.71s &  4.85s &  1.02s &  1.08s\\
Heart 		 &23.52s &	23.26s & 96.08s  & 174.19s	& 188.32s  	& 3.76s  & 4.11s & 75.65s & N/A & N/A &  1.75s &  1.95s\\
Defect 		 & 6.38s &	6.39s & 145.05s  & 336.04s  	& 349.23s  	& 6.92s  & 8.38s & 83.66s & N/A  & N/A   &  2.71s &  3.33s\\
Bank 			&6.39s & 6.45s & 223.03s & 436.02s  	& 458.97s  	& 9.39s  & 10.95s & 151.14s&  N/A  &  N/A  &  4.24s &  4.16s\\
PenDigits 	& 5.53s &  5.47s & 62.12s	 & 71.76s  	& 72.28s  	& 2.17s  & 2.56s & 49.68s & 5.70s &  7.53s & 1.13s  &  1.31s\\
Ailerons 	& 6.53s &	6.56s & N/A    & 370.20s  	&392.15s  	& 7.73s  & 10.08s & N/A&  N/A  & N/A  &  2.96s &  3.97s\\
Credit 		&5.31s &  5.20s & 5.99s		 & 17.51s  	& 20.89s  	& 1.52s  & 1.76s & 5.55s  & 2.66s & 4.52s &  1.00s &  1.08s\\
Satellite 	& 4.03s & 5.30s & 14.49s 	     & 30.24s  	& 36.08s  	& 1.63s  & 1.91s & 10.57s & 3.50s & 5.40s &  1.05s &  1.20s\\
Elevators 	&6.33s &	6.24s & 289.13s  & 427.75s	& 442.75s  	& 8.39s  & 10.48s & 149.87s& N/A & N/A &  3.34s &  4.07s\\
Telescope 	& 4.85s &	4.88s & 181.88s  & 7.19s 	& 9.20s  	& 1.33s  & 1.57s & 108.58s& 1.94s & 3.78s &  0.89s &  1.04s\\
MFeat 	 	& 41.08s &	41.57s & 147.99s  & 301.05s  	& 315.66s  	& 6.81s  & 9.53s & 169.62s  & N/A & N/A & 2.77s &  3.86s\\ \hline\hline 
\textbf{AVG.} & 9.73s &  9.89s & 105.73s   & 183.35s   & 193.43s    & 4.37s  & 5.41s & 72.62s & 3.40s & 5.28s  & 1.97s  & 2.35s \\ \hline\hline
\textbf{SilentWood's Speedup}  & 2.2x & 1.8x & 25.8x  & , 41.9x   & 35.7x   & --  & -- & 38.6x & 3.3x & 4.6x  & --  & -- \\ \hline\hline
\bottomrule
\end{tabular}

\caption{Comparing inference time over various data bit-lengths (14 bits, 16 bits, and 32 bits) w/o GPU support. N/A denotes GPU (or RAM) failure due to large-memory-consuming FHE parameters.}
\label{tab:result-generic}

\end{table*}

%% file: 60-evaluation.tex
\begin{table}[t]
\centering

\begin{minipage}[t]{0.49\linewidth}
\vspace{0pt}
\centering

\setlength{\tabcolsep}{2.0pt} 
\tiny 

\begin{tabular}{|l||r|r|r|r|}
\toprule
\textbf{Bench.}
  & {\textbf{Base}} & \textbf{Base+\textsf{B}} & \textbf{Base+\textsf{B+K}} &  \textbf{Base+\textsf{B+K+C}} \\
&&&&\textbf{(SilentWood)}\\
\hline
Spam & 162.53s  & 9.89s & 6.59s &5.26s \\
Steel & 38.32s & 3.06s & 2.26s &1.98s\\
Breast & 28.27s & 2.84s & 2.21s &1.81s\\
Heart & 188.32s & 8.58s & 4.06s &4.11s\\
Defect & 349.23s & 15.68s & 9.43s &8.38s\\
Bank & 458.97s & 16.20s & 12.20s &10.95s\\
Pendigits & 72.28s & 3.66s & 2.86s &2.56s\\
Ailerons & 392.15s & 19.74s & 12.24s &10.08s\\
Credit & 20.89s & 2.18s & 1.79s &1.76s \\
Satellite & 36.08s & 2.52s & 2.27s &1.91s \\
Elevators & 442.75s & 21.62s & 11.37s &10.48s \\
Telescope & 9.20s & 1.92s & 1.74s &1.57s \\
MFeat & 315.66s & 22.49s & 13.29s &9.53s \\\hline\hline
\textbf{AVG.} & 193.43s & 10.03s & 6.33s & 5.41s \\\hline\hline
\multicolumn{2}{|l|}{\textbf{\shortstack[l]{Speedup against Baseline}}}
  & 19.3$\times$ & 30.6$\times$ & 35.7$\times$ \\
\bottomrule
\end{tabular}%

\captionof{table}{Ablation study.}
\label{tab:speedup-factors}

\end{minipage}
\hfill
\begin{minipage}[t]{0.49\linewidth}
\vspace{0pt}
\centering

\setlength{\tabcolsep}{2.0pt} 
\tiny 
\renewcommand{\arraystretch}{1.2}

\begin{tabular}{|c||cccccccc|}
\hline
& \textbf{$\bm s=0$} & \textbf{$0.01$} & \textbf{$0.05$} & \textbf{$0.1$}
& \textbf{$0.2$} & \textbf{$0.4$} & \textbf{$0.6$} & \textbf{$1$} \\\hline\hline
{Spam} & 0.94 & 0.93 & 0.93 & 0.93 & 0.93 & 0.92 & 0.92 & 0.92   \\
{Steel} & 1.00 & 1.00 & 1.00 & 1.00 & 1.00 & 1.00 & 1.00 & 1.00   \\
{Breast} & 0.67 & 0.67 & 0.66 & 0.66 & 0.66 & 0.64 & 0.59 & 0.59   \\
{Heart} & 0.74 & 0.8 & 0.81 & 0.81 & 0.81 & 0.81 & 0.81 & 0.81   \\
{Defect} & 0.82 & 0.82 & 0.82 & 0.82 & 0.8 & 0.81 & 0.81 & 0.81   \\
{Bank} & 0.99 & 0.99 & 0.99 & 0.99 & 0.99 & 0.99 & 0.99 & 0.99   \\
{PenDigits} & 1.00 & 1.00 & 0.99 & 0.99 & 0.99 & 0.99 & 0.99 & 0.99  \\
{Ailerons} & 0.88 & 0.88 & 0.88 & 0.88 & 0.88 & 0.88 & 0.88 & 0.88   \\
{Credit} & 1.00 & 1.00 & 1.00 & 1.00 & 1.00 & 1.00 & 1.00 & 1.00  \\
{Satellite} & 0.99 & 0.99 & 0.99 & 0.99 & 0.99 & 0.99 & 0.99 & 0.99  \\
{Elevators} & 0.8 & 0.8 & 0.81 & 0.82 & 0.8 & 0.81 & 0.81 & 0.81   \\
{Telescope} & 1.00 & 1.00 & 1.00 & 1.00 & 1.00 & 1.00 & 1.00 & 1.00  \\
{MFeat} & 0.94 & 0.94 & 0.94 & 0.94 & 0.94 & 0.94 & 0.94 & 0.94  \\\hline\hline
\textbf{AVG.} & 0.90 & 0.90 & 0.90 & 0.90 & 0.90 & 0.90 & 0.90 & 0.90  \\\hline
\end{tabular}%

\captionof{table}{Model accuracy at clustering intensities $\bm s$}
\label{tab:result-node-clustering-accuracy}

\end{minipage}

\end{table}

\para{Setup:} We implemented \sysname in C++ (5518 lines) and Python (737 lines). We used the SEAL library~\cite{seal} for CPU-based FHE, and  extended it with PhantomFHE~\cite{phantomfhe} for GPU-based FHE. Our experiments used an i9-14900K 24-core CPU, 168GB RAM, and an RTX 4090 GPU. To train and evaluate XGBoost, we used 13 UCI datasets~\cite{uci}, whose numbers of features, classes, and sample sizes are summarized in \autoref{tab:benchmark-spec} (\autoref{appendix:node-clustering}). We trained the model based on the dataset ratio of training : validation : testing as 6:2:2. 
We compared the performance of the four systems: XGBoost based on Zama's Concrete ML (based on the TFHE-rs library)~\cite{concrete-ml}, SoK-GGG (an MPC protocol based on a 2-party garbled circuit)~\cite{knlas19}, Baseline (RCC-PDTE~\cite{rcc-pdte} with MultiplyPath~\cite{cdpp22}), and \sysname.
The default round-trip time (RTT) was 50ms, the default number of trees was 100, the default maximum tree depth was 7, and the default intensity $s$ of node clustering was 0.2. We measured inference time, inference accuracy, and ciphertext size. We also conducted an ablation study for our proposed techniques over various tree counts (100, 200, 300), maximum tree depths (7, 11, 15), and RTTs (0 ms, 50ms, 100 ms, 200ms) by using the netem tool.

\para{Summary of Results:} For CPU-FHE, \sysname's end-to-end inference time (50ms RTT) is faster than Baseline by 35.7x, faster than Zama by 25.8x, and faster than SoK-GGG by 2.2x. \sysname is faster than SoK-GGG on average even when RTT is 0 ms, and the speedup grows to 2.94x when RTT is 200 ms. Compared to Baseline, \sysname reduces the server-client communication size to roughly 1/5 on average. For GPU-FHE, \sysname is faster than Baseline by 4.6x and faster than Zama by 38.6x.

\para{End-to-end Inference Time: } We evaluated end-to-end inference time with an RTT of 50ms (including the server and client's encoding/decoding, encryption/decryption, and homomorphic operations) for the cases where the feature and threshold bit sizes are 16 bits and 32 bits (quantized), each. Results are illustrated in \autoref{tab:result-generic}. We tested only 14 bits for Zama because its framework supports a maximum 14 bit data length. We evaluated w/o GPU. Without a GPU, Baseline's average computation time is 183.35 s (16 bits) and 193.43 s (32 bits), while Zama's average time is 105.73 s (14 bits). In contrast, \sysname's time is 4.37 s (16 bits) and 5.41 s (32 bits). When using a GPU, Baseline's average computation time is 3.40s (16 bits) and 5.28s (32 bits), and Zama's average time is 72.62s (14 bits), whereas \sysname's average time is 1.97s (16 bits) and 2.35s (32 bits). The FHE-based systems (Zama, Baseline, \sysname) required more runtime for the 32-bit data size, whereas SoK-GGG (MPC) was less-sensitive to the data bit-length but more sensitive to tree depth. For some datasets, SoK-GGG's runtime is shorter for 32-bit than for 16-bit because the trained 32-bit model's tree depths were shallower.

\para{Ablation Study: } \autoref{tab:speedup-factors} illustrates the speedup contribution of \sysname's each technique by evaluating the following versions: Baseline plus BCC (denoted as Baseline\textsf{ + B}), plus computation clustering (denoted as Baseline\textsf{ + B + K}), and plus ciphertext compression (denoted as Baseline\textsf{ + B + K + C}). We used CPU-FHE and an RTT of 50ms for these evaluations. Regarding the speedup factors, BCC's contribution was the largest (19.3x); computation clustering was next (1.58x), and ciphertext compression was last (about 1.17x). Note that the speeedup by ciphertext compression grows as the RTT increases from 50ms to 100ms and 200ms (\autoref{tab:result-rtt}).

\pgfplotstableread[col sep=comma]{data.csv}\datatable

\para{Accuracy of Node Clustering:} 
\autoref{tab:result-node-clustering-accuracy} shows that the average model accuracy across all benchmarks (measured by the testing datasets) does not drop below 0.90 even with high clustering intensity $s$ values, because after each round of node clustering, we validate (by using the validation datasets) the inference accuracy of the updated XGBoost model and abort the change if the model accuracy decreases. This way, we preserve model correctness even under aggressive clustering.

\begin{table*}[t]
\centering
\setlength{\tabcolsep}{3.5pt}
\renewcommand\arraystretch{1.2}

{
\scriptsize

\centering
\begin{tabular}{|l||rrrr|rrrr|rrrr|}
\toprule
\multirow{2}{*}{
{\textbf{Benchmark}}}
  & \multicolumn{4}{c|}{{\textbf{SoK-GGG}}} & \multicolumn{4}{c|}{{ \textbf{Baseline+\textsf{B+K}}}} & \multicolumn{4}{c|}{{ \textbf{Baseline+\textsf{B+K+C}}}} \\
\cmidrule{2-5} \cmidrule{6-9} \cmidrule{10-13}
&\textsf{0ms}&\textsf{50ms}&\textsf{100ms}&\textsf{200ms}&\textsf{0ms}&\textsf{50ms}&\textsf{100ms}&\textsf{200ms}&\textsf{0ms}&\textsf{50ms}&\textsf{100ms}&\textsf{200ms}\\
\hline
Spam & 3.35s & 5.95s & 8.47s &13.44s& 5.92s & 6.59s & 7.92s &10.11s& 4.57s & 5.26s & 6.06s&7.70s\\
Steel & 3.22s & 5.56s & 7.47s &11.91s& 1.23s & 2.26s & 3.43s &5.57s& 1.19s & 1.98s & 2.70s&4.31s\\
Breast & 2.28s & 5.69s & 5.62s &9.16s& 0.94s & 2.21s & 3.05s &5.30s& 0.79s & 1.81s & 2.40s&3.94s\\
Heart & 13.69s & 23.26s & 33.09s &52.88s& 3.52s & 4.06s & 5.71s &7.92s& 3.39s & 4.11s & 4.72s&6.34s\\
Defect & 3.48s & 6.39s & 9.25s &15.06s& 8.97s & 9.43s & 10.95s &13.27s& 7.72s & 8.38s & 8.93s&10.48s\\
Bank & 3.50s & 6.45s & 9.42s &15.96s& 11.95s & 12.20s & 14.08s &16.25s& 10.23s & 10.95s & 11.51s&13.00s\\
Pendigits & 3.42s & 5.47s & 7.77s &12.49s& 1.69s & 2.86s & 3.91s &6.08s& 1.81s & 2.56s & 3.23s&4.74s\\
Ailerons & 3.49s & 6.56s & 9.30s &16.15s& 12.18s & 12.24s & 14.24s &16.36s& 9.37s & 10.08s & 10.66s&12.17s\\
Credit & 3.09s & 5.20s & 6.91s &11.84s& 0.92s & 1.79s & 3.08s &5.30s& 1.02s & 1.76s & 2.43s&3.93s\\
Satellite & 3.04s & 5.30s & 7.54s &12.32s& 1.05s & 2.27s & 3.29s &5.43s& 1.11s & 1.91s & 2.61s&4.19s\\
Elevators & 3.46s & 6.24s & 9.30s &15.21s& 11.57s & 11.37s & 13.57s &15.64s& 9.76s & 10.48s & 10.96s&12.38s\\
Telescope & 3.16s & 4.88s & 6.53s &9.94s& 0.58s & 1.74s & 2.81s &5.06s& 0.83s & 1.57s & 2.25s&3.73s\\
MFeat & 24.20s & 41.57s  & 58.61s &95.23s& 11.33s & 13.29s & 14.9s &15.76s& 8.70s & 9.53s & 10.17s&12.22s\\\hline\hline
\textbf{AVG.} & 5.64s & 9.88s & 13.79s &22.43s& 5.53s & 6.33s & 7.76s &9.85s& 4.65s & 5.41s & 6.05s&7.63s\\\hline
\textbf{SilentWood's Speedup} & 1.21x & 1.83x &2.28x  &2.94x& 1.19x & 1.17x & 1.28x &1.29x& --&--&--&--\\
\bottomrule
\end{tabular}
}
\caption{End-to-end server-client inference time over various RTTs (32-bit feature size)}
\label{tab:result-rtt}
\end{table*}

\para{Inference Time over Various RTTs: } \autoref{tab:result-rtt} shows XGBoost's end-to-end inference time for various RTTs. In the case of 50ms RTT, the time difference between Baseline\textsf{ + B + K} (6.33s) and Baseline\textsf{ + B + K + C} (5.41s) implies the 1.19x speedup achieved by ciphertext compression. \sysname is also faster than SoK-GGG by 1.83x, 2.28x, and 2.94x for RTT 50ms, 100ms, and 200ms respectively. \sysname is faster than SoK-GGG by 1.21x on average even for RTT = 0ms, illustrating \sysname's computational efficiency.

\para{Additional Evaluations:} \autoref{appendix:more-exp} provides the following experiment results: performance comparison across different tree counts and depths (\autoref{appendix:various-models}), the overhead of network, CPU, and memory across different modes (\autoref{appendix:overhead}), varying the number of node clusters, the validation set size, and the applicability of node clustering to MPC (\autoref{appendix:node-clustering}), and inference time across different network bandwidths (\autoref{appendix:bandwidth}).

%% file: 80-related.tex
\para{FHE vs 2PC for PDTE:} 
RCC-PDTE~\cite{rcc-pdte} uses CW-encoding-based comparison to improve this limitation, yet its SumPath aggregation is incompatible with standard forest inference and becomes costly with many trees. \cite{bpsw07} and \cite{knlas19} design constant-round 2PC protocols for PDTE from oblivious transfer and garbled circuits (GC), but their communication grows exponentially with tree depth for evaluating all paths; oblivious RAM (\cite{tkk19}) or pseudo-random functions (\cite{bsccr22}) help only modestly. Gazelle-style FHE+GC hybrids~\cite{217515} inherit this cost for tree traversal (as SoK-GGG does), whereas BCC's only interactive step exchanges packed FHE ciphertexts.

\para{Scaling PDTE to Random Forests:} HBDT (\cite{hbdt}) uses FHE with one-hot encoded features, but one-hot inflates ciphertext count (e.g., a 32-bit feature needs 655 ciphertexts). BPDTE-CW (\cite{cw-batching}) uses one non-zero digit per node to batch comparisons, yet communication still grows (e.g., +17.7\% when precision increases from 11 to 16 bits). Moreover, since summing path scores (SumPath) dominates cost, faster comparisons alone yield limited speedup. Both HBDT and BPDTE-CW aggregate results via MultiplyPath, multiplying along decision paths and incurring heavy ciphertext multiplications. In contrast, \sysname uses summation with BCC to avoid multiplications. 
Kangaroo (\cite{kangaroo}) combines packed HE and secret sharing for amortized private inference over large tree ensembles in WANs. Compared with \sysname, it assumes a weaker threat model (semi-honest server and client), reveals an obfuscated tree topology that constrains the original tree structure, and requires 4 RTTs. \sysname supports malicious-server privacy, hides tree topology from the client, and uses 2 RTTs.

\para{Other Efficiency Protocols:} 
SortingHat (\cite{cdpp22}) and Onion-ORAM (\cite{onion-oram}) are based on TFHE, much slower than \sysname because they lack SIMD.
SortingHat’s transciphering accounts for ~35\% of inference time and grows with tree depth; Onion-ORAM’s ciphertext expansion adds 33–66\% overhead (see Table 2 in their paper). \sysname’s decompression overhead is only 4.5\% (\autoref{tab:result-compression-bcc-overhead}), independent of depth, making it a more efficient compression method for BFV/BGV. 

\para{Tree Compression:} Various pruning techniques reduce inference cost by trimming nodes or trees, e.g., \cite{forestprune}, \cite{lop}, (\cite{fipe}. In contrast, our node clustering preserves the full model structure and merges similar thresholds per feature to reduce distinct comparisons while preserving logic structure. This approach is beneficial when preserving the tree structure is important for interpretability or certification workflows. \sysname’s node clustering complements pruning methods by unifying similar threshold values even after pruning a model.

%% file: 90-conclusion.tex
\sysname's private inference over gradient-boosting forests significantly enhances performance compared to existing methods. \sysname's novel techniques include node clustering to eliminate duplicate homomorphic computations, blind code conversion to preserve arithmetic compatibility for efficient tree score aggregation, and ciphertext compression to reduce the encrypted data size. \sysname achieves an average private XGBoost inference time of 1.8x $\sim$ 41.9x faster than state-of-the-art FHE or MPC-based protocols (i.e., Zama's concrete ML, RCC-PDTE, and SoK-GGG).

%% file: 100-appendix.tex
\section{Operations of the BFV Scheme}
\label{appendix:bfv-scheme} 

The inputs and outputs of the BFV scheme's operations are defined as follows:

\begin{enumerate}
\setlength{\itemsep}{0.5em}
\setlength{\parsep}{0pt}

\item
$\textsf{Encode}\bm{(}[v_0, v_1, \cdots, v_{n-1}]\bm{)}$
\textcolor{red-brown}{$\rhd$ each $v_i \in \mathbb{Z}_t$ for plaintext modulus $t$}\\
$\rightarrow M(X)$
\textcolor{red-brown}{$\rhd$ $M(X)\in \mathbb{Z}_t[X]/(X^n+1)$}

\item
$\textsf{Decode}\bm{(}M(X)\bm{)} \rightarrow [v_0, v_1, \cdots, v_{n-1}]$

\item
$\textsf{Encrypt}\bm{(}M(X), pk\bm{)}$
$\rightarrow ct$
\textcolor{red-brown}{$\rhd$ a ciphertext that encrypts $\Delta \cdot M(X)$ under $sk = S(X)\in\{-1,0,1\}^n \subset \mathbb{Z}[X]/(X^n+1)$}

\item
$\textsf{Decrypt}\bm{(}ct, sk\bm{)} \rightarrow M(X)$

\item
$\textsf{PAdd}\bm{(}ct_1, M_2(X)\bm{)} \rightarrow ct$
\textcolor{red-brown}{$\rhd$ encrypts $\Delta\cdot\bm(M_1(X)+M_2(X)\bm)$, which decodes to $\textsf{Decode}\bm{(}M_1(X)\bm{)} + \textsf{Decode}\bm{(}M_2(X)\bm{)}$}

\item
$\textsf{CAdd}\bm{(}ct_1, ct_2\bm{)} \rightarrow ct$
\textcolor{red-brown}{$\rhd$ encrypts $\Delta\cdot\bm(M_1(X)+M_2(X)\bm)$, which decodes to $\textsf{Decode}\bm{(}M_1(X)\bm{)} + \textsf{Decode}\bm{(}M_2(X)\bm{)}$}

\item
$\textsf{PMult}\bm{(}ct_1, M_2(X)\bm{)} \rightarrow ct$
\textcolor{red-brown}{$\rhd$ encrypts $\Delta\cdot\bm(M_1(X)\cdot M_2(X)\bm)$, which decodes to $\textsf{Decode}\bm{(}M_1(X)\bm{)} \odot \textsf{Decode}\bm{(}M_2(X)\bm{)}$ (where $\odot$ is element-wise vector multiplication)}

\item
$\textsf{CMult}\bm{(}ct_1, ct_2\bm{)} \rightarrow ct$
\textcolor{red-brown}{$\rhd$ encrypts $\Delta\cdot\bm(M_1(X)\cdot M_2(X)\bm)$, which decodes to $\textsf{Decode}\bm{(}M_1(X)\bm{)} \odot \textsf{Decode}\bm{(}M_2(X)\bm{)}$ \\(assuming \textsf{CMult} includes modulus switching to rescale $\Delta^2 \rightarrow \Delta$)}

\item
$\textsf{Rotate}\bm{(}ct, r\bm{)} \rightarrow ct'$
\textcolor{red-brown}{$\rhd$ if $\textsf{Decode}\bm(M(X)\bm)=[v_0,\ldots,v_{n-1}]$, then $\textsf{Decode}\bm(M'(X)\bm)=[v_r,\ldots,v_{n-1},v_0,\ldots,v_{r-1}]$}
\end{enumerate}

\clearpage

\section{Detailed Composition of Private Decision Tree Evaluation (PDTE)}
\label{appendix:pdte-fhe} 

\subsection{Value Comparison: Homomorphic Logic Gates}
\label{appendix:logic-comparison} 

As explained in \autoref{subsec:fhe}, the only arithmetic operations directly supported by FHE are addition and multiplication ($+, \cdot$). Therefore, to homomorphically compare two numbers, one needs to design an arithmetic logic circuit comprising only ($+, \cdot$) operators such that the gate returns 1 if $\alpha > \beta$, and returns false if $\alpha \leq \beta$.

Given the input feature $X$ node and threshold $Y$ are $n$-bits long, one way to build a value comparison circuit is to encrypt each binary bit of $X$ and $Y$ as $(x_1, x_2, \cdots, x_n)$ and $(y_1, y_2, \cdots, y_n)$ and design the value comparator logic as follows:

\begin{itemize}
\setlength{\itemsep}{0.5em}
\setlength{\parsep}{0pt}

\item $\textsf{\textbf{EQBit}}(a, b) = 1 - a - b \pmod 2$ \textcolor{red-brown}{ $\rhd$ bit $a$ is equal to bit $b$}

\item $\textsf{\textbf{GTBit}}(a, b) = a\cdot (1 - b) \pmod 2$ \textcolor{red-brown}{ $\rhd$ bit $a$ $>$ $b$}

\item $\textsf{\textbf{EQ}}_{i\rightarrow j}(X, Y) = \prod\limits_{j=i}^{k}\textsf{EQBit}(x_j, y_j)$ \textcolor{red-brown}{ $\rhd$ Are $X,Y$'s bits $i\sim k$ equal?}

\item $\textsf{\textbf{GT}}(X, Y) = \sum\limits_{i=0}^{n-1}\bm(\textsf{EQ}_{i+1\rightarrow n-1}(X, Y) \cdot \textsf{GTBit}(x_i, y_i)\bm)$ \textcolor{red-brown}{ $\rhd$ $X > Y ?$}

\end{itemize}

Given the above setup, \textsf{GT}$(X, Y)$ returns 1 if $X > Y$; 0 otherwise. Since the above formulas comprise only addition and multiplication operations, they can be executed by using FHE. However, using a bit-wise comparator requires the later tree path evaluation operation to be performed also as a verbose binary circuit, which is not an optimal design for FHE. This is because binary circuits usually involve a deep multiplication chain and FHE requires an increasingly more computation time for a greater multiplication depth.

\subsection{Edge Formulas for SumPath and MultiplyPath}
\label{appendix:edge-formulas} 

\begin{figure}[h]
\centering
{\includegraphics[width=0.6\linewidth]{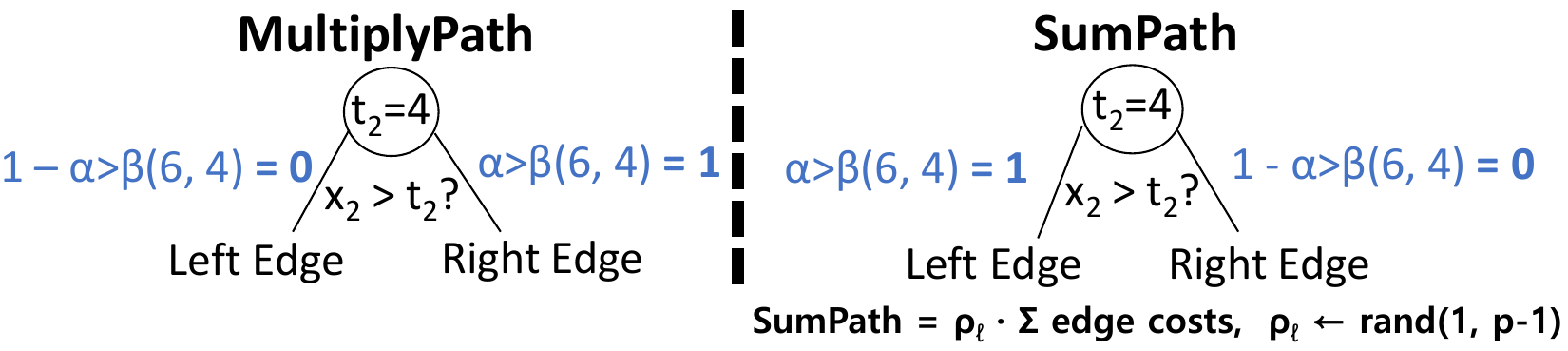}}
\caption{Edge formulas for SumPath and MultiplyPath}
\label{fig:path-formula}
\end{figure}

\autoref{fig:path-formula} shows the per-edge formulas. Let $b_i = \alpha{>}\beta(x_i, t_i) \in \{0,1\}$ be the encrypted comparison result at node $i$. MultiplyPath assigns the indicator $b_i$ to the edge taken when the comparison holds and $1-b_i$ to the other edge, and multiplies the indicators along each root-to-leaf path. SumPath assigns the complementary \emph{cost} $1-b_i$ to the edge taken when the comparison holds and $b_i$ to the other edge, and sums the costs along each path. The cost of leaf $j$ is therefore
\begin{equation}
\label{eq:path-cost}
c_j = \sum_{i \in \mathit{path}(j)} \mathit{cost}_i \;\in\; \{0, 1, \ldots, d_j\},
\end{equation}
which is the number of unsatisfied edges on the path, where $d_j \le d$ is the depth of leaf $j$. $c_j = 0$ iff $j$ is the true leaf of its tree, and otherwise $c_j \ge 1$ because at least the edge where the path departs from the true path is unsatisfied. Computing $c_j$ requires only \textsf{CAdd} (and a plaintext negation for $1-b_i$). To hide $c_j$ from the client, the server multiplies each leaf cost by a fresh mask $\rho_j \xleftarrow{\$} \mathbb{Z}_p^{*}$, so the SumPath value of leaf $j$ is $\rho_j c_j$. In \sysname, this mask is applied once to the merged ciphertext (Algorithm~\autoref{alg:blind-shuffling}, line 10), so path evaluation costs one plaintext multiplication per ciphertext rather than one per node.

Two properties follow because $p$ is prime (a requirement of BFV batching anyway) and $d < p$. First, no false path can vanish: $c_j \in \{1, \ldots, d\} \subseteq \mathbb{Z}_p^{*}$ and $\rho_j \in \mathbb{Z}_p^{*}$, and a field has no zero divisors, so $\rho_j c_j \neq 0$. Second, $\rho_j c_j$ is \emph{uniform over $\mathbb{Z}_p^{*}$} and independent across leaves, since $\rho \mapsto \rho c_j$ is a bijection on $\mathbb{Z}_p^{*}$ and the $\rho_j$ values are independent.

\subsection{Value Comparison: CW-encoding}
\label{appendix:cw-comparison}

\begin{figure}[h]
\centering
{\includegraphics[width=0.6\linewidth]{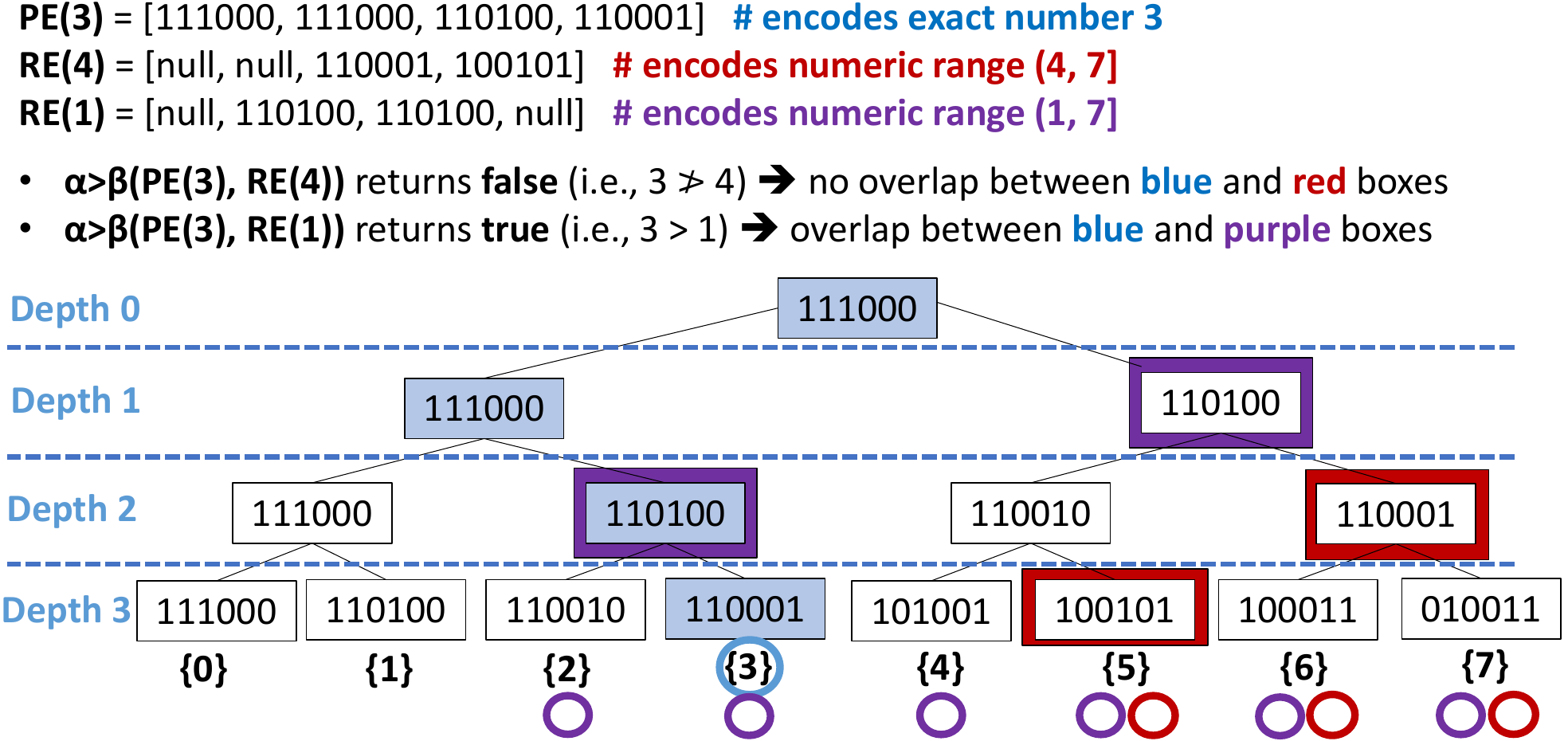}}
\caption{RCC-PDTE~\cite{rcc-pdte}'s examples of value comparison ($3>4?$  \textit{ }  $3>1?$), where each tree node's label is CW-encoded.}
\label{fig:example-comparison}
\end{figure}

\begin{figure}[h]
\centering
{\includegraphics[width=0.5\linewidth]{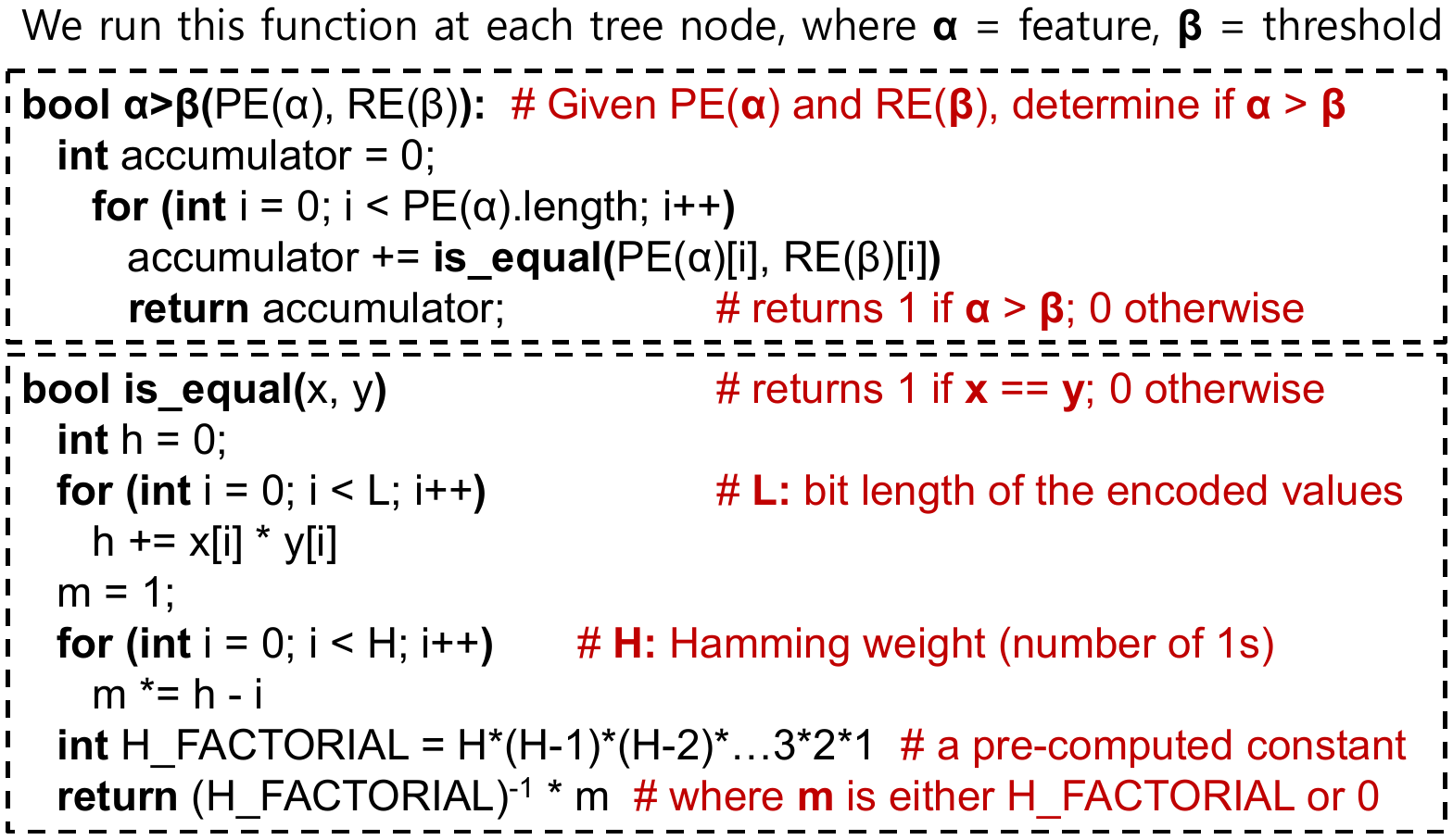}}
\caption{An arithmetic circuit of comparing two input numbers ($\alpha, \beta$) encoded by CW-encoding (PE($\alpha$), RE($\beta$)).}
\label{fig:cw-algorithm}
\end{figure}

The state-of-the-art value comparison algorithm based on FHE is the one proposed by RCC-PDTE~\cite{rcc-pdte}. \autoref{fig:example-comparison} illustrates a high-level idea of how this works: two numbers to be compared are encoded as a vector of node labels that form a complete binary tree (CBT, which is different from a decision tree). CBT's each leaf node corresponds to the target number to encode. The CBT has 8 leaves and can encode any numbers between 0 and 7. Based on this CBT, RCC-PDTE proposes two number encoding schemes: PE (point encoding) and RE (range cover encoding). PE encodes an exact number corresponding to a particular leaf node in a CBT as a vector of node labels. The elements of this vector tell the path from the root to a target leaf node. In \autoref{fig:example-comparison}, we PE-encode 3 as PE vector $[111000, 111000, 110100, 110001]$ (visualized as blue nodes and a blue circle). The CBT's each node label is encoded as constant-weight encoding~\cite{cw-encoding} (CW-encoding), which is essentially a Hamming-weight encoding whose base-2 representation always contains the same frequency of 1s for any target value to encode. We will later explain why we use CW-encoding for labeling. 

Unlike PE encoding, RE encodes a range of values as a vector such that each of its elements corresponds to a specific node in each tree depth (or can be \textsf{NULL} if no node is chosen at that depth), and the encoded range of values is the set of all descendant leaves of the nodes in the vector. In \autoref{fig:example-comparison}, we RE-encode range $(4, 7]$ as RE vector $\mathit{RE}(4) = [\mathit{null}, \mathit{null}, 110001, 100101]$ (visualized as red nodes), whose each element's descendant leaves collectively cover value range $\{5, 6, 7\}$ (red circle). As another example, we RE-encode range $(1, 7]$ as $\mathit{RE}(1) = [\mathit{null}, 110100, 110100, \mathit{null}]$ (purple nodes), whose descendant leaves collectively cover value range $\{2, 3, 4, 5, 6, 7\}$ (purple circles). Notably, the upper bound of the range encoded by RE is always the right-most leaf value in CBT (e.g., 7 in case of \autoref{fig:example-comparison}). 

Based on this CBT setup, to evaluate the condition $\alpha > \beta$, we determine whether two vectors $\mathit{PE}(\alpha)$ and $\mathit{RE}(\beta)$ share any common node (i.e., same node label at the same depth of CBT). If yes, then $\alpha > \beta$ holds; otherwise $\alpha \leq \beta$. For example, $\mathit{PE}(3)$ and $\mathit{RE}(4)$ have no common node at any depth, which visually implies that the blue path of $\mathit{PE}(3)$ has no overlap with the red boundary of $\mathit{RE}(4)$ (i.e., $3 \leq 4$). On the other hand, $\mathit{PE}(3)$ and $\mathit{RE}(1)$ have a common node at depth 2 (110100), which visually implies that the blue path of $PE(3)$ has an overlap with the purple boundary of $RE(1)$ (i.e., $3 > 1$). This value comparison logic is implemented in the upper dashed box of \autoref{fig:cw-algorithm}. Within this function, the for-loop statement can be executed with FHE as a series of $(+, \cdot)$ operations, and we further detail the \textsf{is\_equal} function's internal logic in the lower dashed box, which again comprises only $(+, \cdot)$ operations. Therefore, the entire value comparison operation can be performed by FHE with the target numbers being encrypted ciphertexts. In particular, \textsf{is\_equal} assumes that the CBT's each node label is encoded as CW-encoding. Note that all for-loops in these functions have a deterministic number of looping, because the conditional variables for looping (\textsf{.length}, \textsf{L}, and \textsf{H}) are statically known values.
The benefit of using CW encoding in this comparison algorithm is that during the FHE computation, we can effectively fix the number of multiplicative depth (fixed to $\lceil\log H\rceil$ + 2) required for the \textsf{is\_equal} operation regardless of the bit-length of input data ($x, y$).

\subsection{Tree Path Evaluation: MultiplyPath and SumPath}
\label{appendix:multiplypath-sumpath}

\begin{figure*}[h]
\centering
{\includegraphics[width=1.0\linewidth]{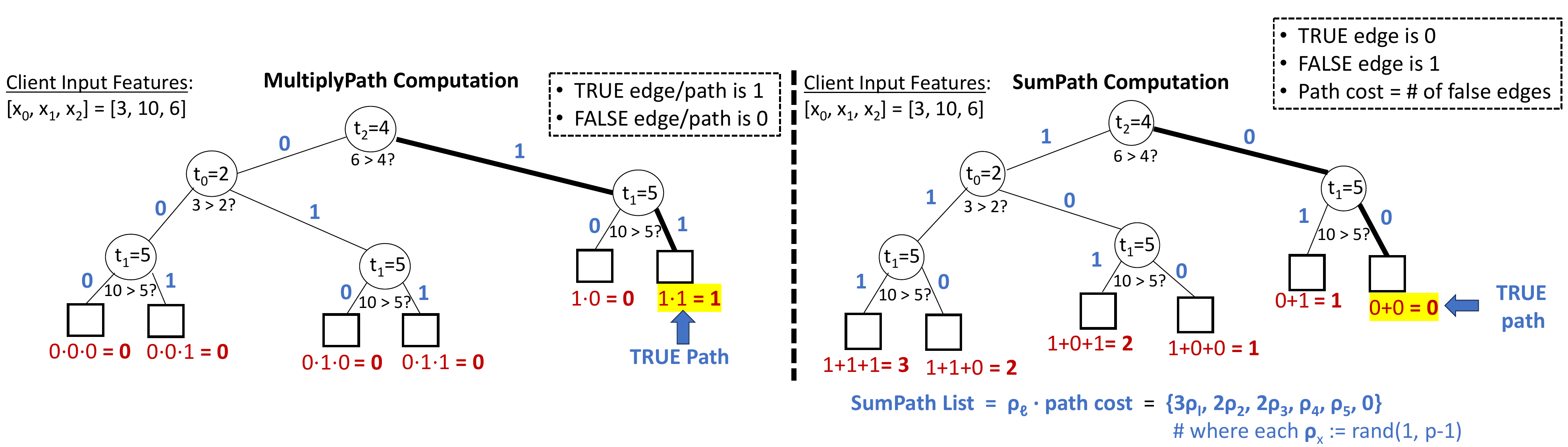}}
\caption{A tree path evaluation example of MultiplyPath and SumPath}
\label{fig:path-computation}
\end{figure*}

In \autoref{fig:path-computation}, the client's input is $[x_0, x_1, x_2] = [3, 10, 6]$, so every comparison holds ($6 > 4$, $3 > 2$, $10 > 5$) and, at every node, the left edge (taken when the comparison fails) has cost $1$ while the right edge has cost $0$. For MultiplyPath the indicators are complementary, and only the rightmost leaf has product $1$. For SumPath, the six leaves (left to right) obtain the costs $3, 2, 2, 1, 1, 0$; only the rightmost leaf, the true path, is $0$. After masking with per-leaf $\rho_\ell$, the SumPath values are $3\rho_1, 2\rho_2, 2\rho_3, \rho_4, \rho_5, 0$: all non-zero except the true path, and each non-zero value is uniform over $\mathbb{Z}_p^{*}$.

\clearpage

\section{An Example and Overhead Analysis of the Blind Code Conversion Protocol}

\subsection{An example of the Blind Code Conversion (BCC) Protocol}
\label{appendix:bcc-example}

\begin{figure}[h]
\centering
{\includegraphics[width=0.6\linewidth]{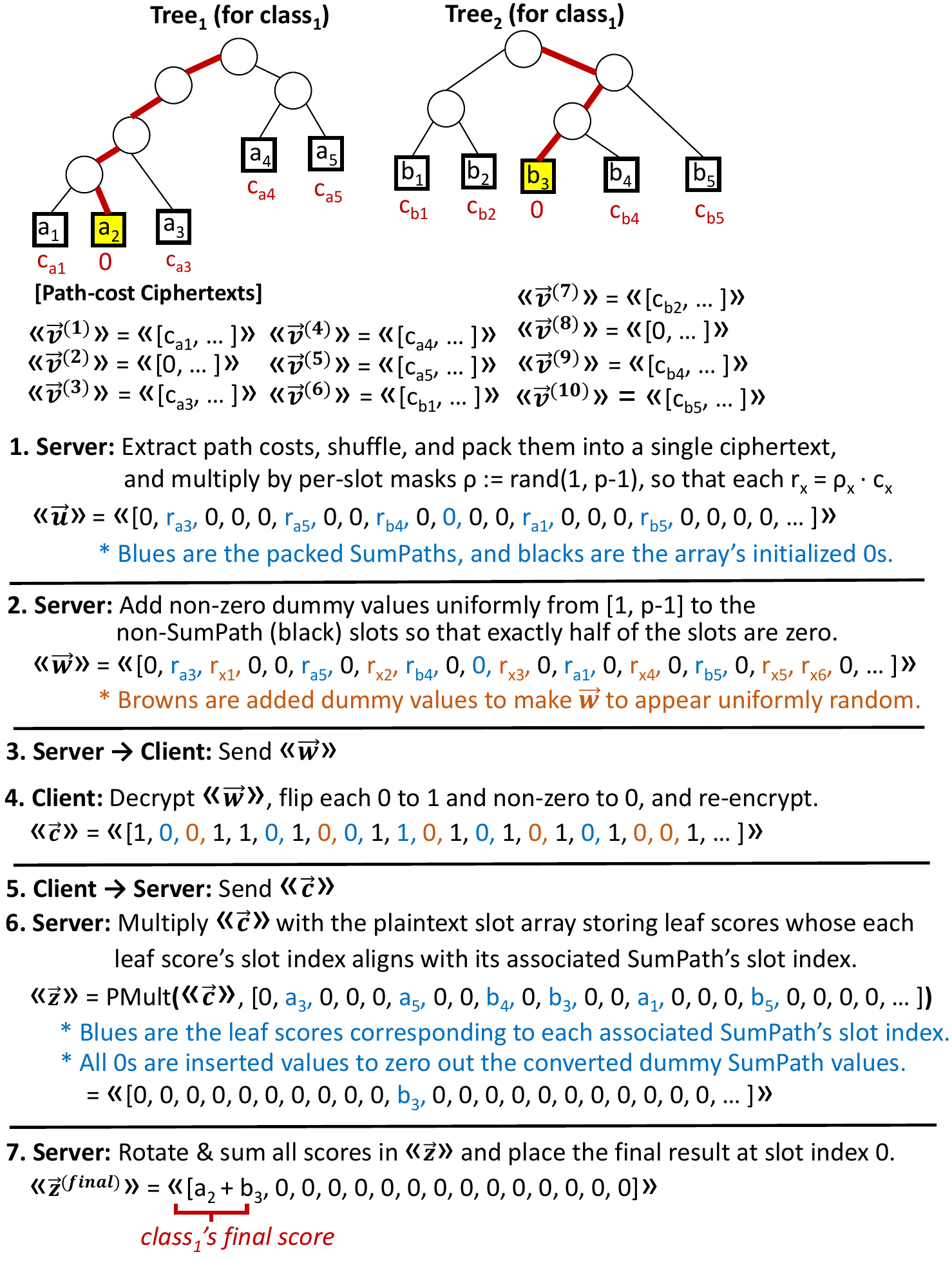}}
\caption{An example of BCC for XGBoost scoring.}
\label{fig:bcc-example}
\end{figure}

\autoref{fig:bcc-example} shows an example of XGBoost score computation based on the BCC protocol. 
Initially, the server has a list of path-cost ciphertexts, each of which contains a distinct (unmasked) path cost $c_i$ (\autoref{eq:path-cost}) in slot index 0 and zero values in other slots. 
In step 1, the server extracts only the encrypted path costs, randomly rotates each of them, merges them into a new array $\vec{u}$, and multiplies the merged ciphertext by a per-slot mask $\vec{\rho}$ whose entries are uniform over $\mathbb{Z}_p^{*}$ (one \textsf{PMult}). The resulting slot values $r_x = \rho_x c_x$ are the SumPaths shown in the figure. The output of step 1 is $\llangle \vec{u} \rrangle$ (where $\llangle \rrangle$ denotes encryption). 
In step 2, the server homomorphically adds dummy values to $\vec{u}$ (non-zero dummies drawn uniformly from $\mathbb{Z}_p^{*}$) to make the ratio between the zeros and non-zeros uniform (1:1 in \autoref{fig:bcc-example}'s example) and randomly ordered. This ratio is a fixed hyperparameter of the protocol. 
In step 3, the server sends the ciphertext to the client. 
In step 4, the client decrypts it, flips 0s to 1s and non-zeros to 0s, re-encrypts them, and sends them back to the server. During this process, the client learns nothing about the real SumPath values: since $p$ is prime and $1 \le c_j \le d < p$, every false-path value $\rho_j c_j$ is non-zero and uniform over $\mathbb{Z}_p^{*}$, so the decrypted slot array is distributed exactly as $f_{\mathit{zero}}$ zeros at random positions and $N - f_{\mathit{zero}}$ i.i.d. $\mathbb{Z}_p^{*}$-uniform values, whatever the model and input (Lemma~\ref{lem:view}). In particular, exactly $T$ of the zeros are true paths, and no false path is ever converted to $1$. The server also learns nothing about the SumPath values because they always remain encrypted when handled by the server. 
In step 5, the server receives the ciphertext containing client-converted values. 
In step 6, the server homomorphically multiplies the ciphertext by the server's plaintext array that encodes the leaf values (scores), where each array slot's leaf value aligns with its associated SumPath's slot index. Thus, when we multiply this array with the client-returned ciphertext, the leaf values for false SumPaths and dummy values are masked to 0, while those for true SumPaths remain. 
In step 7, the server sums all per-tree scores according to the class by rotating and adding the slot values to compute the final score for each class. 
Finally, the server compactly packs all final class scores into $\llangle \vec{z}^{(\mathit{final})} \rrangle$ and sends them to the client. In the case of binary classification, a single final score is returned to the client, and whether this score is positive or negative indicates the binary classification result. In the case of multi-class classification, the highest class score is regarded as the inferred class.

In \autoref{fig:bcc-example}, steps $1\sim 5$ comprise the generic BCC protocol (which we formally describe in \autoref{tab:bcc}), whereas steps $6\sim 7$ involve application-specific score computation for XGBoost.

\subsection{Analysis of the Score Aggregation Overhead}
\label{appendix:bcc-complexity}

\begin{table}[h]
\setlength{\tabcolsep}{5pt}
\renewcommand{\arraystretch}{1.2}
  \scriptsize
  \centering  
  \begin{tabular}{|c||c|c|}  
    \hline
    & \textbf{BCC} & \textbf{MultiplyPath}\\\hline\hline
    \textsf{CMult} & 0 &  $O(\ell)$ \\\hline
\textsf{PMult} & $O(\log N) + 1 \approx O(1)$ & $O(\ell)$ \\\hline
\textsf{Rotate} & $O(\ell + \log N)  \approx O(\ell)$ & $0$ \\\hline
    \textsf{CAdd} & $O(\ell + \log N) \approx O(\ell)$ & $O(\ell)$ \\\hline
    \text{RTTs} & 1 & 0 \\\hline
  \end{tabular}
\caption{Comparing the runtime overhead of leaf score aggregation based on BCC \textit{vs.} MultiplyPath.}  
  \label{tab:bcc-complexity}
\end{table}

\begin{table}[h]
\setlength{\tabcolsep}{5pt}
\renewcommand{\arraystretch}{1.2}
  \scriptsize
  \centering  
  \begin{tabular}{|c||c|c|}  
    \hline
    & \textbf{BCC} & \textbf{MultiplyPath}\\\hline\hline
    \textsf{CMult} & 0 &  $O(d)$ \\\hline
\textsf{PMult} & $O(\log N) + 1 \approx O(1)$ & $O(1)$ \\\hline
\textsf{Rotate} & $O(\log N) \approx O(1)$ & $0$ \\\hline
    \textsf{CAdd} & $O(\log \ell + \log N)$ & $O(\log \ell)$ \\\hline
  \end{tabular}
\caption{Circuit depth (i.e., critical path) of each type of operation.}  
  \label{tab:bcc-critical-path}
\end{table}

$N$ is the number of plaintext slots in a ciphertext, $\ell$ is the total number of tree paths, and $d$ is the tree depth (i.e., path length). In this time complexity analysis, we assume a complete binary tree. Therefore, there are a total of $2\ell - 1$ nodes. In practical FHE, $\log N$ is fixed to be between 13 and 15; thus, $O(\log N) \approx O(1)$. 

\autoref{tab:bcc-complexity} shows the time complexity of score aggregation based on BCC and MultiplyPath. 
As illustrated in Algorithm~\autoref{alg:blind-shuffling}, score aggregation based on BCC requires $\ell$ rotations and $\ell-1$ additions to randomly permute and aggregate $\ell$ real path costs (lines 6$\sim$9), 1 plaintext multiplication to mask them (line 10), and 1 addition to merge them with randomly permuted dummy SumPaths (line 14). 
The overhead of $O(\log N)$ rotations and $O(\log N)$ plaintext-ciphertext multiplications occur during score aggregation (line 6, 7, and 8 in Algorithm~\autoref{alg:blind-scoring}), but this overhead is fixed and small in practice because $\log N$ is fixed to be between 13 and 15. 
Additionally, BCC involves one round-trip communication between the server and the client. During score aggregation in Algorithm~\autoref{alg:blind-scoring}, BCC requires $O(\log N) \approx O(1)$ ciphertext-to-plaintext multiplications to aggregate the final score (line 8).  
The masking \textsf{PMult} does not add to the cost of path evaluation either: the original SumPath formulation multiplies every node's comparison result by a random value (one \textsf{PMult} per node) before summation, whereas \sysname sums the $0/1$ edge costs and applies a single \textsf{PMult} per merged ciphertext afterwards. In both cases the multiplicative depth after the comparison circuit is one plaintext multiplication with plaintext magnitude bounded by $p$, so the noise budget is unchanged.

In contrast, for a full binary tree with $\ell$ leaves, there are
$\ell-1$ internal nodes and $2\ell-2$ edges. With shared path
prefixes, MultiplyPath computes all leaf indicators using
$O(\ell)$ \textsf{CMult}s. Weighting the $\ell$ leaf indicators
requires $\ell$ \textsf{PMult}s, followed by $\ell-1$
\textsf{CAdd}s to aggregate the scores. Since all these
operations are performed locally by the server, no round-trip
communication with the client is needed. However, the dominant cost arises from the $O(\ell)$ \textsf{CMult} operations, which is the most computationally expensive and consumes the largest noise budget among all (i.e., it is generally 10 times slower than \textsf{PMult} or \textsf{Rotate}).

\autoref{tab:bcc-critical-path} shows the circuit depth (i.e., the critical path) of score aggregation based on BCC and MultiplyPath. 

The $\ell$ rotations are mutually independent.
Combining their outputs requires $\ell-1$ ciphertext additions,
with depth $\lceil\log_2\ell\rceil$ when implemented as a
balanced reduction. 
The $\ell$ rotations are mutually independent. Combining their outputs requires $\ell-1$ ciphertext additions, with depth $\lceil\log_2\ell\rceil$ when implemented as a balanced reduction. The subsequent score aggregation contributes another $\log_2N$ sequential levels.

With multi-core processors or GPUs, this for-loop can also run efficiently in parallel. On the other hand, the critical path of MultiplyPath is $d$ \textsf{CMult} operations during path evaluation. This is a significant computational bottleneck because, as the tree depth grows linearly, the computational overhead grows polynomially.

\clearpage

\section{Security Analysis: Blind Code Conversion}
\label{appendix:blind-shuffling-analysis}

Our protocol achieves privacy against both semi-honest clients and malicious servers.\footnote{Privacy against malicious clients can be achieved by attaching a proof of well-formedness to all keys and ciphertexts.} 
In particular, the client only learns the final class scores and some hyperparameters of the server's model, 
while the server learns essentially nothing about the client's features.
Below we formally prove security with respect to each type of corruption when the protocol is instantiated with a sanitized FHE scheme.

The idealized functionality $\mathcal{F}_{\text{BCC}}$ of the protocol is defined as follows:
\begin{itemize}
    \item The client chooses its input for the classification.
    \item The server chooses its decision forest model, including some hyperparameters such as the frequency distribution $f$ and the upper bound on the number of SumPaths $\ell_{\mathsf{max}}$.
    \item The client obtains the final class scores computed from the decision forest and the hyperparameters $f$ and $\ell_{\mathsf{max}}$.
\end{itemize}

\para{Correctness:}
Let $p$ be the (prime) plaintext modulus and $d < p$ the maximum tree depth. For every leaf $j$, the path cost satisfies $c_j \in \{0, \ldots, d\}$ with $c_j = 0$ iff $j$ is the true leaf of its tree (\autoref{eq:path-cost}). Since $\mathbb{Z}_p$ is a field and $\rho_j \in \mathbb{Z}_p^{*}$, the masked value $\rho_j c_j$ is $0$ iff $c_j = 0$. Hence the conversion table $\mathcal{U}_p$ maps exactly the $T$ true-path slots of $\vec{w}$ to $1$ (all dummy slots being either $0$ or $\mathbb{Z}_p^{*}$-valued), Algorithm~\autoref{alg:blind-scoring} sums exactly the true-path leaf scores, and the number of zeros in $\vec{w}$ is exactly $f_{\mathit{zero}}$. The protocol therefore has no probabilistic decoding error,
provided that homomorphic evaluation decrypts correctly. Assigning independent random values to individual edges and summing them instead would let a false path with $k \ge 2$ unsatisfied edges vanish with probability $\approx 1/p$, i.e., $\approx \ell/p$ per query.

\begin{lemma}[Distribution of the client's intermediate view]
\label{lem:view}
Let $p$ be prime and $d < p$; let $\vec{\rho} \xleftarrow{\$} (\mathbb{Z}_p^{*})^{N}$ and the non-zero dummies be sampled independently and uniformly from $\mathbb{Z}_p^{*}$; and let $\vec{\sigma}$ be a uniformly random permutation with $f_{\mathit{zero}}$ zero slots. 
We assume $T \leq f_{\mathit{zero}}$ and $\ell - T \leq N - f_{\mathit{zero}}$. 
Then, for every model $y$ and every input $x$, the plaintext $m_1 = \vec{w}$ underlying $c_1$ is distributed as follows: the positions of the $f_{\mathit{zero}}$ zero slots form a uniformly random subset of $[N]$, and the remaining $N - f_{\mathit{zero}}$ slots are i.i.d. uniform over $\mathbb{Z}_p^{*}$.
\end{lemma}
\begin{proof}
Each true path contributes a $0$, and the $f_{\mathit{zero}} - T$ dummy zeros contribute the rest, so there are exactly $f_{\mathit{zero}}$ zeros. For a false path $j$, $c_j$ is a fixed element of $\{1, \ldots, d\} \subseteq \mathbb{Z}_p^{*}$, and $\rho \mapsto \rho c_j$ is a bijection on $\mathbb{Z}_p^{*}$, so $\rho_j c_j$ is uniform over $\mathbb{Z}_p^{*}$; independence across paths follows from the independence of the $\rho_j$. The non-zero dummies are i.i.d. uniform over $\mathbb{Z}_p^{*}$ by construction. Finally, $\vec{\sigma}$ places all values at uniformly random positions. None of these distributions depends on $y$ or $x$.
\end{proof}

\para{Security Against Corrupted Client:}
We demonstrate how the client's view may be simulated using only its own input, the hyperparameters and the final class scores.
The client's view consists of its own input, its random coins, and the transcript of the protocol~\cite{DBLP:books/sp/17/Lindell17}.
Formally, we have
$$\mathsf{view}_{\mathsf{Clnt}}(x, y, \lambda) = (x, r, c_1, c_2),$$
where $x$ holds the client's private features, $y$ is the server's private model, $\lambda$ is the security parameter, $r$ is the collection of the client's random coins, $c_1$ is the ciphertext holding the padded and shuffled SumPath values, and $c_2$ is the ciphertext holding the final class scores.

We require a sanitization algorithm, which transforms a ciphertext in a manner that preserves the message, while removing all information about the evaluated circuit. In other words, when sanitizing two ciphertexts that decrypt to the same message, the outputs are statistically close (hence indistinguishable).
We refer to Ducas and Stehlé~\cite{DBLP:conf/eurocrypt/DucasS16} for a formal treatment of sanitization.
Assuming that $c_1$ and $c_2$ are sanitized, we can build a simulator for the client's view as follows:
\begin{itemize}
    \item The simulator copies the value of $x$ from its input.
    \item The simulator generates $r$ uniformly at random.
    \item By Lemma~\ref{lem:view}, the distribution of the plaintext $m_1$ underlying $c_1$ depends only on the public $f$ and $\ell_{\mathsf{max}}$, not on $x$, $y$, or the other entries of the client's view (it is masked, padded, and shuffled using the server's random coins). Therefore, the simulator samples $m_1$ exactly from this distribution ($f_{\mathit{zero}}$ zeros at uniformly random positions, i.i.d. $\mathbb{Z}_p^{*}$-uniform values elsewhere), so the simulated $m_1$ is identically distributed to the real one. It then generates the encryption key from the client's random coins, and it encrypts and sanitizes $m_1$. Due to the sanitization property, the resulting ciphertext is indistinguishable from the actual $c_1$.
    \item The simulator encrypts and sanitizes the final class scores. Due to the sanitization property, the resulting ciphertext is indistinguishable from the actual $c_2$.
\end{itemize}

\para{Security Against Corrupted Server:}
\sysname can be modeled as a client-aided outsourcing protocol. Akavia et al.~\cite{DBLP:journals/joc/AkaviaGHV25} show that such a protocol achieves privacy if at least one of the following conditions is satisfied:
\begin{itemize}
    \item It is cleartext computable (all plaintexts/cleartexts are functions of the client's input) and it is instantiated with a CPA-secure FHE scheme.
    \item It is instantiated with a funcCPA-secure FHE scheme.
\end{itemize}
The first condition is not satisfied in \sysname, because the plaintexts depend on the server's model and its padding \& random shuffling.
However, Akavia et al.~\cite{DBLP:journals/joc/AkaviaGHV25} show that sanitized CPA-secure schemes are funcCPA-secure.
Therefore, security can be proven in one of these situations: (i) BFV is funcCPA-secure; or (ii) BFV is CPA-secure and all ciphertexts are sanitized before sending them to the server.

\clearpage

\section{Homomorphic Ciphertext Decompression with Example}
\label{appendix:ciphertext-decompression}

\begin{figure}[h]
\centering
{\includegraphics[width=0.5\linewidth]{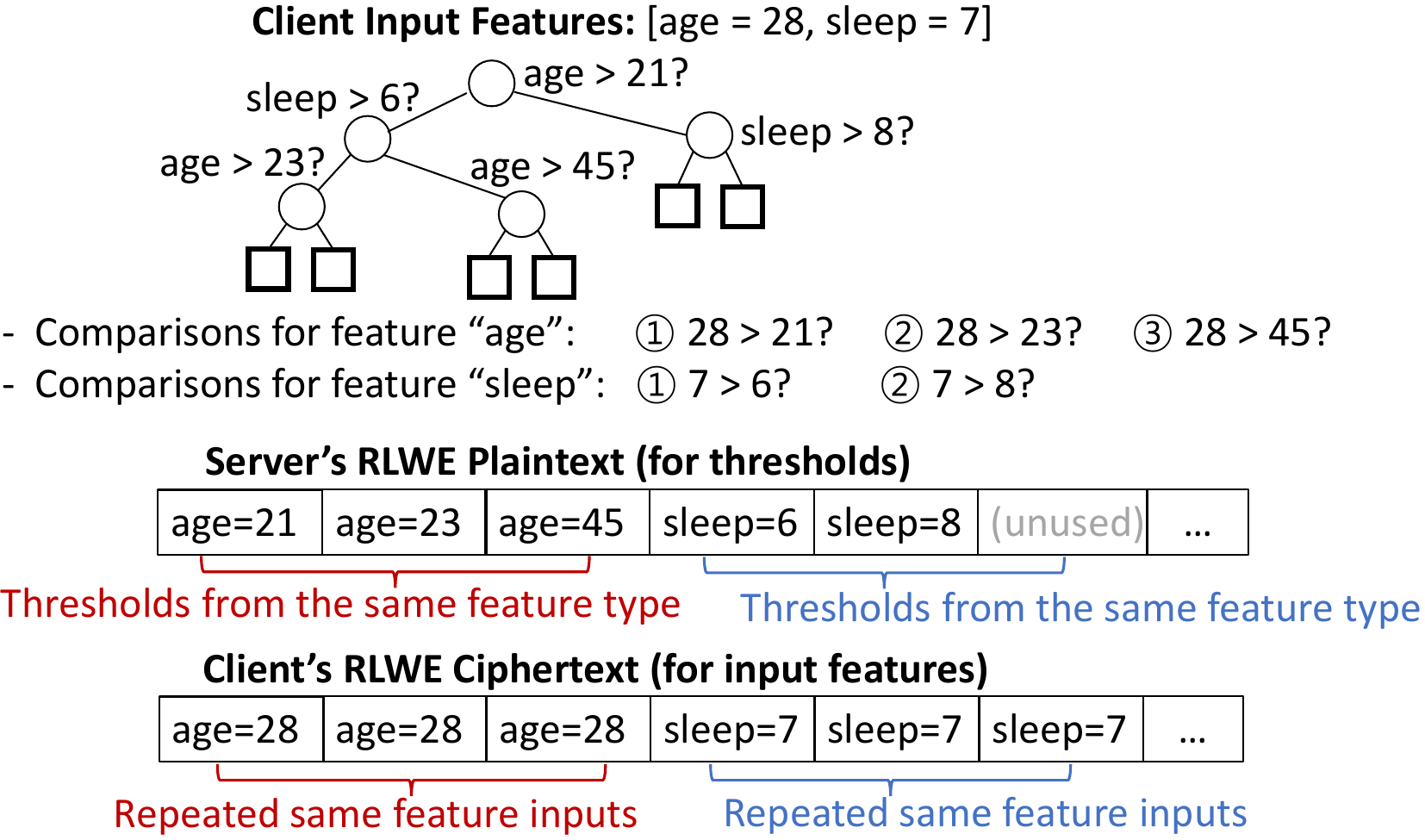}}
\caption{An example of repetitive data encoding}
\label{fig:repetitive-data-encoding}
\end{figure}

\begin{figure}[h]
\centering
{\includegraphics[width=1.0\linewidth]{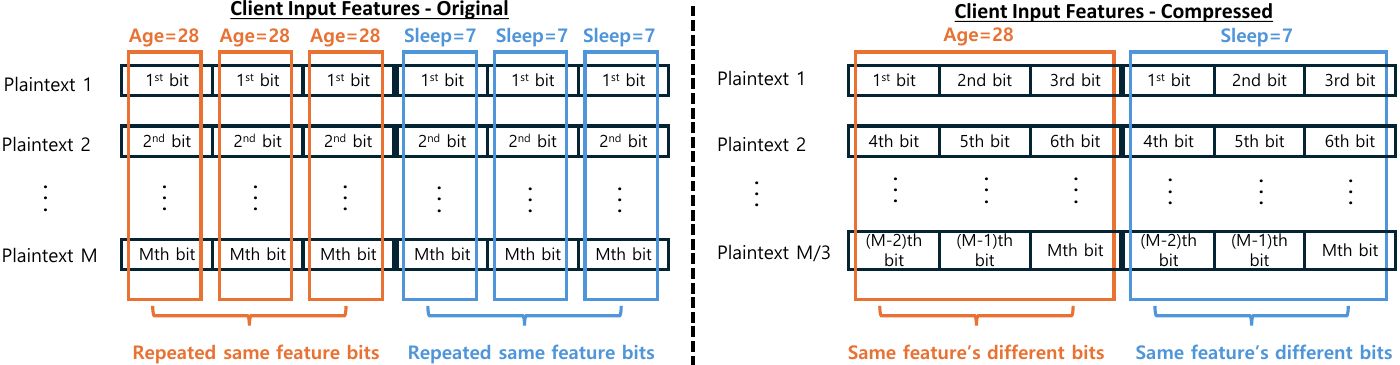}}
\caption{An example of ciphertext compression}
\label{fig:ciphertext-compression}
\end{figure}

\autoref{fig:repetitive-data-encoding} is an example showing how PDTE generally uses repetitive data encoding: the server's tree has 3 nodes conditioned on the \textit{age} feature (each threshold is 21, 23, and 45) and 2 nodes conditioned on the \textit{sleep} feature (each threshold is 6 and 8). The server homomorphically compares these threshold values against the client's encrypted \textit{age=28} and \textit{sleep=7} input feature values in a SIMD manner. Since the server has 3 distinct threshold values for \textit{age} and 2 distinct threshold values for \textit{sleep}, the client encodes its \textit{age=28} and \textit{sleep=7} values repetitively 3 times (capping to the server's maximum number of comparisons for any same feature name). While such repetitive data encoding enables efficient SIMD comparisons, this repetition wastes ciphertext slot utility.

\autoref{fig:ciphertext-compression} shows the resulting example of \sysname's ciphertext compression. In FHE-based private protocols, when a client CW-encodes feature data, s/he has to create as many ciphertexts as the number of feature data bits ($M$ in \autoref{fig:ciphertext-compression}'s example), where each ciphertext repetitively encodes each feature's same bit 3 times (to help the server-side SIMD operation). On the other hand, using \sysname's ciphertext compression, the client removes these repetitions of data before encrypting them, which results in 2 empty slots for each feature bit storage in the ciphertext. Therefore, the client can store the bits of other digits of data in this available extra space. Applying this compact data encoding, the total required number of ciphertexts reduces from $M$ to $\frac{M}{3}$. Once the server receives these compressed ciphertexts, it homomorphically performs a reverse operation of what the client did, effectively decompressing and restoring the originally intended data encoding (i.e., 3 repetitions of feature bits). 

\clearpage

\begin{algorithm}[h]
\scriptsize
\begin{algorithmic}[1]
\STATE {\textbf{Input:} $\{C_0,\ldots,C_{K-1}\}$, $\{\mathit{start}_j\}$, $\{\mathit{gap}_0,\ldots,\mathit{gap}_{R-1}\}$, $R$ \\\textcolor{red-brown}{$\rhd$ $K=\lfloor M/R\rfloor$ compressed ciphertexts; same $\mathit{start}_j$, $\mathit{gap}_r$, $R$ as in \autoref{alg:ciphertext-compression}}}
\FOR{$k \in \{0, \ldots, K-1\}$}
    \FOR{$r \in \{0, \ldots, R-1\}$ \textcolor{red-brown}{$\rhd$ recover $A_{kR+r}$}}
        \STATE {$\vec{m} \gets [0,\ldots,0]$;\; $m[\mathit{start}_j + \mathit{gap}_r] \gets 1$ \textbf{for all} $j$ \textcolor{red-brown}{$\rhd$ mask selecting the $r$-th repetition slots}}
        \STATE {$E \gets \textsf{PMult}(C_k, \textsf{Encode}(\vec{m}))$ \textcolor{red-brown}{$\rhd$ only $A_{kR+r}$'s bits remain, at slots $\mathit{start}_j+\mathit{gap}_r$}}
        \STATE {$D_{kR+r} \gets \sum_{r'=0}^{R-1} \textsf{Rotate}(E,\; \mathit{gap}_{r'} - \mathit{gap}_r)$ \textcolor{red-brown}{$\rhd$ copy them to all $R$ repetition slots ($\textsf{CAdd}$; the $r'{=}r$ term is $E$ itself)}}
    \ENDFOR
\ENDFOR
\STATE {\textbf{Output:} $\{D_0, \ldots, D_{KR-1}\}$ \textcolor{red-brown}{$\rhd$ $D_i$ = $A_i$ with every bit repeated $R$ times, exactly as repetitive encoding}}
\end{algorithmic}
\caption{Ciphertext Decompression}
\label{alg:ciphertext-decompression}
\end{algorithm}

Algorithm~\autoref{alg:ciphertext-decompression} reverses Algorithm~\autoref{alg:ciphertext-compression} on the server side: for each compressed ciphertext $C_k$ and each $r$, it masks the slots of the $r$-th repetition (one \textsf{PMult}) and copies them into the other $R-1$ repetition slots with $R-1$ rotations and additions, so that $D_{kR+r}$ encrypts exactly what the client would have sent under repetitive encoding of $A_{kR+r}$.

\FloatBarrier

\clearpage

\section{Additional Evaluations}
\label{appendix:more-exp}

\subsection{\sysname's Performance Across Various Model Sizes}
\label{appendix:various-models}

\begin{table}[h]
\setlength{\tabcolsep}{3.0pt}
\scriptsize
\renewcommand\arraystretch{1.3}
\centering
\begin{tabular}{|l||rrr|rrr||rrr|rrr|}
\toprule
\multirow{2}{*}{
{\textbf{Bench.}}}
  & \multicolumn{3}{c|}{{\textbf{Inference Time}}} &
\multicolumn{3}{c||}{ \textbf{Model Accuracy}} & \multicolumn{3}{c|}{{ \textbf{Inference Time}}} & \multicolumn{3}{c|}{{ \textbf {Model Accuracy}}} \\ 
\cmidrule{2-4} \cmidrule{5-7} \cmidrule{8-10} \cmidrule{11-13}
&\textsf{100T}&\textsf{200T}&\textsf{300T}&\textsf{100T}&\textsf{200T}&\textsf{300T}&\textsf{7D}&\textsf{11D}&\textsf{15D}&\textsf{7D}&\textsf{11D}&\textsf{15D}\\
\hline
Spam & 3.82s & 5.49s & 6.68s & 0.94 & 0.94 & 0.94 & 3.82s & 4.42s & 4.92s & 0.94 & 0.94 & 0.94 \\
Steel & 1.81s & 2.33s & 2.69s &  1.00 & 1.00 & 1.00 & 1.81s & 1.79s & 1.73s &  1.00 & 1.00 & 1.00  \\
Breast & 1.57s & 1.69s & 1.82s &  0.67 & 0.67 & 0.67 & 1.57s & 1.45s & 1.46s &  0.67 & 0.67 & 0.67  \\
Heart & 3.76s & 4.97s & 6.67s &  0.74 & 0.74 & 0.74 & 3.76s & 3.18s & 3.86s &  0.74 & 0.74 & 0.74  \\
Defect & 6.92s & 12.73s & 14.39s &  0.79 & 0.81 & 0.80 & 6.92s & 11.23s & 16.38s &  0.79 & 0.81 & 0.80  \\
Bank & 9.39s & 10.66s & 12.42s & 0.99 & 0.99 & 0.99  & 9.39s & 12.47s & 12.51s & 0.99 & 0.99 & 0.99  \\
Pendigits & 2.17s & 2.33s & 2.40s &  0.99 & 1.00 & 1.00 & 2.17s & 2.22s & 2.01s &  0.99 & 1.00 & 1.00 \\
Ailerons & 7.73s & 10.59s & 12.34s & 0.87 & 0.88 & 0.88 & 7.73s & 11.99s & 15.12s & 0.87 & 0.87 & 0.87 \\
Credit & 1.51s & 1.59s & 1.58s & 1.00 & 1.00 & 1.00 & 1.51s & 1.45s & 1.46s & 1.00 & 1.00 & 1.00 \\
Satellite & 1.63s & 1.69s & 1.87s &  0.99 & 0.99 & 0.99 & 1.63s & 1.62s & 1.57s &  0.99 & 0.99 & 0.99  \\
Elevators & 8.39s & 10.82s & 14.49s &   0.80 & 0.80 & 0.80 & 8.39s & 15.38s & 18.02s &  0.80 & 0.80 & 0.80  \\
Telescope & 1.33s & 1.41s & 1.54s &  1.00 & 1.00 & 1.00  & 1.33s & 1.27s & 1.24s &  1.00 & 1.00 & 1.00  \\
MFeat & 6.81s & 6.68s & 7.55s &  0.93 & 0.93 & 0.93 & 6.81s & 6.13s & 6.19s &   0.93 & 0.93 & 0.93 \\\hline\hline
\textbf{AVG.} & 4.37s & 5.61s & 6.65s & 0.90 & 0.90 & 0.90 & 4.37s & 5.74s & 6.65s & 0.90 & 0.90 & 0.90  \\
\bottomrule
\end{tabular}
\caption{Inference time over various numbers of trees (100T$\sim$ 300T) and maximum tree depths (7D$\sim$15D).
}
\label{tab:trees-depths}
\end{table}

\para{Inference Time:} \autoref{tab:trees-depths} shows the average inference time across various numbers of trees (100 $\sim$ 300). The average computation time roughly increased linearly with the number of trees, as their number of rotations to handle SumPath also increased linearly. In \autoref{tab:trees-depths}, we also measured the change in inference time across various maximum tree depths (7, 11, 15). The average inference time increased with depth (not due to BCC, but due to SumPath computation). The increase between 11 and 15 is relatively bigger than that between 7 and 11.

\para{Default Model Accuracy:} \autoref{tab:trees-depths} also illustrates the accuracy of each trained XGBoost model (measured using the testing datasets) across various numbers of trees and maximum tree depths. Note that these measured model accuracies are affected only by the training parameters, not by the optimization techniques of \sysname. The average model accuracy of all benchmarks was 0.9 for all numbers of trees (100$\sim$300) and all maximum tree depths (7$\sim$15). This implies that 100 trees with a maximum depth of 7 are generally a sufficient setup to achieve the best accuracy for private XGBoost inference.

\begin{table}[h]
\centering
\scriptsize
\setlength{\tabcolsep}{5pt}
\caption{Comparison of inference time across various numbers of trees.}
\label{tab:t_runtime}
\begin{tabular}{lrrrrrrrr}
\toprule
& \multicolumn{4}{c}{\textbf{Baseline}}
& \multicolumn{4}{c}{\textbf{SilentWood}} \\
\cmidrule(lr){2-5}
\cmidrule(lr){6-9}
\textbf{Dataset}
& $\mathbf{10\mathsf T}$
& $\mathbf{50\mathsf T}$
& $\mathbf{100\mathsf T}$
& $\mathbf{200\mathsf T}$
& $\mathbf{10\mathsf T}$
& $\mathbf{50\mathsf T}$
& $\mathbf{100\mathsf T}$
& $\mathbf{200\mathsf T}$ \\
\midrule
spam
& 78.50s & 198.29s & 309.42s & 523.18s
& 2.22s & 4.49s & 8.05s & 10.39s \\

steel
& 9.08s & 28.05s & 46.68s & 74.32s
& 1.48s & 1.90s & 2.21s & 2.91s \\

cancer
& 9.09s & 20.69s & 24.82s & 43.12s
& 1.50s & 1.75s & 1.85s & 2.16s \\

heart
& 94.93s & 230.22s & 306.98s & 506.65s
& 2.36s & 5.76s & 6.90s & 9.56s \\

defect
& 118.81s & 449.49s & 833.95s & 1488.26s
& 2.75s & 7.60s & 11.79s & 19.28s \\

bank
& 163.85s & 606.74s & 1183.46s & 1729.59s
& 3.40s & 10.77s & 17.76s & 23.67s \\

pendigits
& 29.35s & 90.93s & 117.17s & 170.98s
& 1.73s & 2.45s & 2.71s & 3.11s \\

ailerons
& 153.87s & 532.60s & 1070.04s & 1537.48s
& 3.35s & 8.23s & 14.68s & 21.71s \\

creditcard
& 5.58s & 7.81s & 15.78s & 27.29s
& 1.45s & 1.59s & 1.75s & 2.11s \\

satellite
& 11.74s & 20.43s & 39.61s & 45.20s
& 1.54s & 1.74s & 2.05s & 2.09s \\

elevators
& 152.44s & 630.26s & 1185.83s & 1683.95s
& 2.97s & 8.95s & 15.01s & 29.89s \\

telescope
& 5.69s & 8.48s & 13.24s & 21.70s
& 1.39s & 1.53s & 1.70s & 2.04s \\

mfeat
& 215.13s & 435.30s & 586.70s & 963.95s
& 5.37s & 10.88s & 12.41s & 15.43s \\

\midrule
\textbf{AVG}
& \text{80.62s}
& \text{250.72s}
& \text{441.05s}
& \text{678.13s}
& \text{2.42s}
& \text{5.20s}
& \text{7.61s}
& \text{11.10s} \\

\textbf{Speedup}
& -- & -- & -- & --
& \textbf{33.26$\mathbf\times$}
& \textbf{48.18$\mathbf\times$}
& \textbf{57.99$\mathbf\times$}
& \textbf{61.08$\mathbf\times$} \\
\bottomrule
\end{tabular}
\end{table}

\para{Inference Time Comparison between Baseline and \sysname}: \autoref{tab:t_runtime} reports the inference time (in seconds) of Baseline and \sysname at RTT=50ms for 10, 50, 100, 200 trees (these supplementary runs were performed on an Intel Ultra 7 255H, 16 Cores). Even when the number of trees is small, \sysname significantly outperforms the Baseline. This is because Baseline's MultiplyPath requires large cryptographic parameters to support long multiplicative chains of homomorphic operations. In contrast, \sysname's path evaluation requires only additive chains, consuming a negligible amount of noise budget. Regardless of the number of trees, the maximum depth of the trees is set as a training hyperparameter, capped at 7 in this evaluation.
As the tree count grows,\sysname's speedup increases because the benefits of our node clustering and BCC become more pronounced.

\subsection{Overhead of Ciphertext Compression and Blind Code Conversion}
\label{appendix:overhead}

\begin{table}[h]
\setlength{\tabcolsep}{3.0pt}
\scriptsize
\renewcommand\arraystretch{1.2}
\centering
\begin{tabular}{|l||rr|rr|rr|}
\toprule
\multirow{2}{*}{
{\textbf{Bench.}}}
& \multicolumn{2}{c|}{\textbf{SoK-GGG}} & \multicolumn{2}{c|}{\textbf{Baseline\textsf{ + B + K}}} & \multicolumn{2}{c|}{\textbf{Baseline\textsf{ + B + K + C}}} \\
\cmidrule{2-3} \cmidrule{4-5} \cmidrule{6-7} &\textsf{16-bit}&\textsf{32-bit}&\textsf{16-bit}&\textsf{32-bit}&\textsf{16-bit}&\textsf{32-bit}\\
\hline
Spam & 181.2MB & 182.4MB & 26.7MB & 85.7MB & 5.8MB & 22.8MB \\
Steel & 35.0MB & 36.3MB & 26.7MB & 85.7MB & 4.5MB & 16.3MB \\
Breast & 18.3MB & 9.8MB & 26.7MB & 85.7MB & 4.5MB & 12.3MB \\
Heart & 219.4MB & 208.8MB & 26.7MB & 85.7MB & 3.2MB & 7.1MB \\
Defect & 324.9MB & 322.5MB & 26.7MB & 85.7MB & 3.4MB & 10.0MB \\
Bank & 401.6MB & 381.2MB & 26.7MB & 85.7MB & 4.7MB & 17.5MB \\
Pendigits & 64.8MB & 66.6MB & 26.7MB & 85.7MB & 3.2MB & 8.4MB \\
Ailerons & 359.4MB & 358.5MB & 26.7MB & 85.7MB & 4.7MB & 16.5MB \\
Credit & 17.6MB & 13.5MB & 26.7MB & 85.7MB & 4.5MB & 12.3MB \\
Satellite & 27.5MB & 30.6MB & 46.4MB & 85.7MB & 4.5MB & 16.3MB \\
Elevators & 401.9MB & 390.8MB & 26.7MB & 85.7MB & 3.4MB & 8.7MB \\
Telescope & 12.6MB & 12.1MB & 26.7MB & 85.7MB & 3.2MB & 5.8MB \\
MFeat & 346.5MB & 341.5MB & 26.7MB & 85.7MB & 7.1MB & 43.8MB \\\hline\hline
\textbf{AVG.} & 185.5MB & 180.8MB &	28.22MB & 85.75MB & 4.34MB & 15.20MB \\
\bottomrule
\end{tabular}
\caption{Communication cost (i.e., exchanged data) between the server and client over various feature bit-lengths}
\label{tab:compression}
\end{table}

\para{Network Overhead of Ciphertext Compression:} \autoref{tab:compression} shows the total size of data exchanged between the server and client during XGBoost inference, where Baseline\textsf{ + B + K + C} and Baseline\textsf{ + B + K} are \sysname w/o ciphertext compression. Generally, \sysname's ciphertext sizes were much smaller than those of SoK-GGG's. For \sysname w/o compression, the exchanged ciphertext size increased with the feature data size. Without the compression technique, the ciphertext sizes remain constant across all datasets. This is mainly because the client's number of ciphertexts for the initial query matched the number of CW-encoded data bits. For BCC and the final inference response, only a single ciphertext was needed, as all intermediate data fit within 16,384 slots of a single ciphertext. By contract, \sysname's ciphertext compression could reduce the number of required ciphertexts by a factor of repetitive data encoding, where the repetition was calculated by dividing the total number of slots in a ciphertext by: the total number of feature types multiplied by the CW-encoded bit-length. On average, using ciphertext compression reduced the ciphertext size from 85.75MB to 15.20MB in the case of 32-bit feature data bit-length. \autoref{tab:result-compression-bcc-overhead} shows the computation overheads for the server's ciphertext decompression (Algorithm~\autoref{alg:ciphertext-decompression} in \autoref{appendix:ciphertext-decompression}); the average times are 142 ms and 241 ms for 16-bit and 32-bit data bit lengths, respectively.

\begin{table}[h]
\scriptsize
\renewcommand\arraystretch{1.2}
\centering
\begin{tabular}{|l||rr|rr|}
\toprule
\multirow{2}{*}{
{\textbf{Bench.}}}
  & \multicolumn{2}{c|}{{ \textbf {Cipher Decompression}}} & \multicolumn{2}{c|}{{ \textbf {BCC Computation}}}  \\ 
\cmidrule{2-3} \cmidrule{4-5} 
&\textsf{16-bit}&\textsf{32-bit} &\textsf{16-bit}&\textsf{32-bit}\\\hline
Spam &  80ms & 102ms&  1608ms & 1629ms \\
Steel &  150ms& 189ms&  717ms & 711ms \\
Breast &   126ms & 207ms&  453ms & 429ms \\
Heart & 205ms & 310ms& 2314ms & 2358ms \\
Defect & 149ms & 250ms&  4248ms & 4454ms \\
Bank &  145ms & 174ms& 4713ms & 4717ms \\
Pendigits&  156ms& 351ms&  833ms & 832ms \\
Ailerons   &  131ms & 180ms&  4223ms & 4495ms \\
Credit  &  111ms & 194ms&  379ms & 380ms \\
Satellite  & 112ms & 178ms& 498ms & 518ms \\
Elevators & 176ms  & 361ms&  5916ms & 6081ms \\
Telescope & 209ms & 412ms &  335ms & 367ms \\
MFeat &  90ms & 224ms &  3433ms & 3733ms \\\hline\hline
\textbf{AVG.}  &  142ms & 241ms& 2282ms & 2362ms \\
\bottomrule
\end{tabular}
\caption{The computation overhead of ciphertext decompression and BCC.}
\label{tab:result-compression-bcc-overhead}
\end{table}

\para{Computation Overhead of Blind Code Conversion:} \autoref{tab:result-compression-bcc-overhead} also shows the computation overheads BCC (\autoref{alg:blind-shuffling} and \autoref{alg:blind-scoring}). A majority of the overhead came from the rotation operations to permute real SumPaths. The average times are 2282ms and 2362ms for 16-bit and 32-bit data bit lengths.

\begin{table}[h]
\centering
\scriptsize
\setlength{\tabcolsep}{6pt}
\caption{Server and client's peak memory usage of Baseline and \sysname{}.}
\label{tab:server_client_cost}
\begin{tabular}{lrrrr}
\toprule
& \multicolumn{2}{c}{\textbf{Baseline}}
& \multicolumn{2}{c}{\textbf{SilentWood}} \\
\cmidrule(lr){2-3}
\cmidrule(lr){4-5}
\textbf{Dataset}
& \textbf{Server} & \textbf{Client}
& \textbf{Server} & \textbf{Client} \\
\midrule
spam       & 15438.8MB & 413.6MB & 3359.8MB & 63.7MB \\
steel      & 6542.3MB  & 413.5MB & 1357.3MB & 58.2MB \\
cancer     & 4543.5MB  & 413.5MB & 1009.6MB & 58.6MB \\
heart      & 23856.0MB & 413.6MB & 4563.3MB & 63.2MB \\
defect     & 28849.3MB & 413.3MB & 6231.1MB & 55.6MB \\
bank       & 29034.2MB & 412.8MB & 7832.8MB & 59.2MB \\
pendigits  & 8547.7MB  & 413.2MB & 1669.4MB & 53.7MB \\
ailerons   & 29029.4MB & 412.7MB & 7594.2MB & 59.8MB \\
creditcard & 4173.5MB  & 413.4MB & 962.3MB  & 58.4MB \\
satellite  & 5604.0MB  & 413.3MB & 1156.8MB & 58.8MB \\
elevators  & 28985.5MB & 412.6MB & 8464.4MB & 55.8MB \\
telescope  & 4072.9MB  & 413.5MB & 935.8MB  & 53.6MB \\
mfeat      & 29082.8MB & 412.9MB & 6887.6MB & 85.8MB \\
\midrule
\textbf{AVG}
& \textbf{16750.8} & \textbf{413.2}
& \textbf{4001.9}  & \textbf{60.3} \\
\bottomrule
\end{tabular}
\end{table}

\para{Peak Memory Usage:} ~\autoref{tab:server_client_cost} measures the peak memory usage of the server and client in MB. \sysname{}'s average peak memory usage is about 4~6 times less than Baseline. The server used the majority of memory for FHE computations.
Both Baseline and \sysname actively run rotations to compute either MultiplyPath or SumPath, as well as to aggregate tree scores, which require significantly more rotations than decompression. Their Galois key sizes are 784.06MB (Baseline) and 195.03MB (\sysname), each.SilentWood's is smaller due to its smaller cryptographic parameters.

\subsection{Further Analysis of Node Clustering}
\label{appendix:node-clustering}

\begin{table*}[h]
\scriptsize
\setlength{\tabcolsep}{5pt}
\renewcommand{\arraystretch}{1.2}
\scriptsize
  \centering  
  \begin{tabular}{|c||c|c|c|c|c|c|c|c|c|c|c|c|c|}  
    \hline
    & Spam & Steel & Breast & Heart  & Defect & Bank & PenDigits & Ailerons & Credit & Satellite & Elevators & Telescope & MFeat \\\hline\hline
    \textbf{Features} & 57 & 33 & 30 & 13 & 32 & 21 & 16 & 40 & 30 & 36 & 18 & 11 & 64 \\\hline
    \textbf{Classes} & 2 & 2 & 2 & 5 & 2 & 2 & 2 & 2 & 2 & 2 & 2 & 2 & 10\\\hline
    \textbf{Samples} & 4,601 & 1,941 & 569 & 294 & 8,191 & 10,885 & 10,992 & 14,750 & 14,143 & 5,100 & 16,599 & 19,020 & 2,000
 \\\hline
  \end{tabular}
\caption{Specification of benchmarks}  
  \label{tab:benchmark-spec}
\end{table*}
\begin{figure}[h]
\centering

{\footnotesize
\resizebox{\linewidth}{!}{%
\begin{tikzpicture}
\begin{axis}[
    ybar,
    bar width=1.8pt,
    x=37pt,
    enlarge x limits=0.05, 
    ymode=log,
    log basis y=10,
    ymin=1,
    ytick={1,10,100,1000,10000},
    height=2.8cm,
    ylabel={\# of clusters},
    xtick=data,
    xticklabels from table={\datatable}{Benchmarks},
    ymajorgrids,
    legend pos=north west,
    legend style={
        at={(0.5,-0.45)},
        anchor=north,
        legend columns=9,
        /tikz/every even column/.style={column sep=0.8em},
        legend image code/.code={
            \draw[##1,draw=none]
                 (0pt,0pt) rectangle (6pt,4pt);
        },
        clip=false
    }
]
   \addplot table [x expr=\coordindex, y=No clustering]{\datatable};
   \addplot table [x expr=\coordindex, y=Exact Match]{\datatable};
   \addplot table [x expr=\coordindex, y=0.01 Range]{\datatable};   
   \addplot table [x expr=\coordindex, y=0.05 Range]{\datatable};   
   \addplot table [x expr=\coordindex, y=0.1 Range]{\datatable};   
   \addplot table [x expr=\coordindex, y=0.2 Range]{\datatable};   
   \addplot table [x expr=\coordindex, y=0.4 Range]{\datatable};   
   \addplot table [x expr=\coordindex, y=0.6 Range]{\datatable};   
   \addplot table [x expr=\coordindex, y=Full Range]{\datatable};   

   \legend{
       No clustering,
       $s=0$ (exact match),
       $s=0.01$,
       $s=0.05$,
       $s=0.1$,
       $s=0.2$,
       $s=0.4$,
       $s=0.6$,
       $s=1.0$
   }
\end{axis}
\end{tikzpicture}%
}
}

\caption{The number of node clusters across various node clustering intensities.}
\label{fig:result-node-clustering}
\end{figure}
\para{Number of Node Clusters:} \autoref{tab:benchmark-spec} summarizes the properties (i.e., features, classes, and sample sizes) of the 13 UCI datasets~\cite{uci} used for \sysname's evaluation. \autoref{fig:result-node-clustering} shows the number of node clusters across various clustering intensities $s$ (0.0 $\sim$ 1.0), where two nodes $n_i$ and $n_j$ are co-clustered if $\frac{|\mathit{n_i.val} - \mathit{n_j.val}|}{\textbf{MAX}(\mathit{n_i.name}) - \textbf{MIN}(\mathit{n_i.name})} < s$. The success of clustering gradually decreased as we increased the intensity $s$. For example, when intensity $s=0.2$, on average, 2,388 nodes in XGBoost were clustered into 46 node clusters (i.e., $\frac{2388}{46} = 51.9$x node reduction rate). While the greater clustering intensity provided a higher opportunity for node clustering, the effectiveness of clustering (i.e., node reduction rate) did not increase further when $s$ was greater than 0.4.

\begin{table*}[t]
\fontsize{7}{8}\selectfont
\setlength{\tabcolsep}{3pt}
\renewcommand{\arraystretch}{1.08}
\centering

\begin{tabular}{|c|l||c|c|c|c|c|c|c|c|c|c|c|c|c|}
\hline
\multirow{2}{*}{\textbf{Val. Set}} & \multirow{2}{*}{\textbf{Metric}} &
\multicolumn{13}{c|}{\textbf{Benchmark}} \\
\cline{3-15}
& & \textbf{Spam} & \textbf{Steel} & \textbf{Breast} & \textbf{Heart} & \textbf{Defect} & \textbf{Bank} &
\textbf{PenDigits} & \textbf{Ailerons} & \textbf{Credit} & \textbf{Satellite} & \textbf{Elevators} & \textbf{Telescope} & \textbf{MFeat} \\
\hline\hline

\multirow{3}{*}{5\%}  & Node Reduction & 71\% & 91\% & 84\% & 98\% & 74\% & 54\% & 79\% & 88\% & 66\% & 81\% & 92\% & 94\% & 76\% \\
                  & Test Accuracy  & 91\% & 100\%& 63\% & 72\% & 61\% & 90\% & 98\% & 83\% & 100\% & 100\%& 74\% & 100\%& 91\% \\
                  & Abortion       & 30\% & 0\%  & 9\%  & 14\% & 70\% & 74\% & 38\% & 52\% & 0\%  & 10\% & 57\% & 0\%  & 27\% \\
\hline

\multirow{3}{*}{10\%} & Node Reduction & 69\% & 91\% & 84\% & 94\% & 79\% & 62\% & 77\% & 89\% & 66\% & 79\% & 90\% & 94\% & 71\% \\
                  & Test Accuracy  & 93\% & 100\%& 61\% & 72\% & 62\% & 91\% & 97\% & 84\% & 100\% & 100\%& 73\% & 100\%& 91\% \\
                  & Abortion       & 40\% & 0\%  & 9\%  & 29\% & 60\% & 61\% & 50\% & 56\% & 0\%  & 13\% & 57\% & 0\%  & 30\% \\
\hline

\multirow{3}{*}{20\%} & Node Reduction & 72\% & 91\% & 78\% & 96\% & 76\% & 68\% & 71\% & 89\% & 66\% & 75\% & 89\% & 94\% & 65\% \\
                  & Test Accuracy  & 95\% & 100\%& 60\%& 81\% & 82\% & 99\% & 99\% & 88\% & 100\% & 99\%& 80\% & 100\%& 94\% \\
                  & Abortion       & 35\% & 0\%  & 23\% & 14\% & 60\% & 55\% & 56\% & 52\% & 0\%  & 26\% & 64\% & 0\%  & 42\% \\
\hline

\multirow{3}{*}{30\%} & Node Reduction & 74\% & 91\% & 72\% & 93\% & 81\% & 62\% & 69\% & 89\% & 62.5\% & 75\% & 90\% & 94\% & 66\% \\
                  & Test Accuracy  & 98\% & 100\%& 60\%& 81\% & 82\% & 93\% & 97\% & 85\% & 100\%   & 100\%& 84\% & 100\%& 89\% \\
                  & Abortion       & 40\% & 0\%  & 23\% & 43\% & 50\% & 68\% & 69\% & 52\% & 9\%    & 26\% & 79\% & 0\%  & 45\% \\
\hline

\multirow{3}{*}{35\%} & Node Reduction & 74\% & 91\% & 70\% & 93\% & 82\% & 62\% & 69\% & 89\% & 62.5\% & 72\% & 90\% & 94\% & 65\% \\
                  & Test Accuracy  & 95\% & 100\%& 80\%& 81\% & 85\% & 93\% & 95\% & 85\% & 100\%   & 100\%& 76\% & 100\%& 91\% \\
                  & Abortion       & 40\% & 0\%  & 27\% & 43\% & 50\% & 65\% & 69\% & 56\% & 9\%    & 35\% & 79\% & 0\%  & 52\% \\
\hline

\end{tabular}
\caption{Node clustering with various validation set sizes (5\%$\sim$35\%).}
\label{tab:node-clustering-more-exp}
\end{table*}

\para{Various Validation Set Sizes:} \autoref{tab:node-clustering-more-exp}  shows model accuracy, clustering efficiency, and merge abortion rate across 
various validation set sizes (5\%$\sim$35\%). A small validation set (5\%) leads to unstable accuracy, which improves and stabilizes as its size increases to 20\%. At 35\%, the test-set size drops too much, causing instability in test accuracy. As the validation-set size increases, clustering efficiency declines, and the abortion rate increases, which is natural. Our default (20\% validation set, 20\% test set) balances these factors effectively.

\begin{table}[h]
\fontsize{7}{8}\selectfont
\setlength{\tabcolsep}{4pt}
\renewcommand{\arraystretch}{1.2} 
  \centering  
  \begin{tabular}{|c||c|c|c|}  
    \hline
    & \textbf{Comparison} & \textbf{Total} & \textbf{Fraction of}\\
    & \textbf{Time} & \textbf{Inference Time} & \textbf{Comparison Time}
    \\\hline\hline
    {Spam} & 5.26s & 6.14s & 85.6\% \\
    {Steel} & 4.46s & 5.39s & 82.7\% \\
    {Breast} & 4.49s & 5.32s & 84.4\% \\
    {Heart} & 20.37s & 24.00s & 84.9\% \\
    {Defect} & 5.55s & 6.41s & 86.6\% \\
    {Bank} & 5.67s & 6.58s & 86.2\% \\
    {PenDigits} & 4.83s & 5.82s & 83.0\% \\
    {Ailerons} & 5.62s & 6.56s & 85.7\% \\
    {Credit} & 4.42s & 5.24s & 84.3\% \\
    {Satellite} & 4.68s & 5.49s & 85.2\% \\
    {Elevators} & 5.56s & 6.49s & 85.7\% \\
    {Telescope} & 4.03s & 5.00s & 80.1\% \\
    {MFeat} & 37.51s & 45.37s & 82.7\% \\\hline\hline
    \textbf{AVG.} & 8.65s & 10.29s & 84.1\% \\\hline
  \end{tabular}
\caption{SoK-GGG's Comparison Times and Total Inference Times}  
  \label{tab:sok-breakdown}
\end{table}

\para{Applicability of Node Clustering to SoK-GGG:} As we showed in \autoref{tab:result-rtt}, SoK-GGG's computation time (without the communication cost) for private inference is large: 5.64 seconds on average. This cost primarily includes the time for garbled circuit operations in the comparison phase: OT extension for input values, circuit garbling (server-side), and circuit evaluation (client-side). \autoref{tab:sok-breakdown} additionally shows SoK-GGG's inference time and the fraction of time spent on comparisons at tree nodes. If we apply node clustering to SoK-GGG, the reduction ratio of nodes linearly reduces the amount of computation spent on the OT extension. 
Let $\alpha$ be the ratio of comparisons before node clustering to comparisons after node clustering, and let $\beta$ be the time spent on comparisons relative to the total time spent in SoK-GGG.  Let $t$ be the time spent in SoK-GGG's garbled circuit for comparisons. Then, SoK-GGG's computation time for comparison with node clustering is expected to run in $\dfrac{(\beta \cdot t)}{\alpha}  + (1 - \beta) \cdot t$.

\begin{table}[h]
\centering
\scriptsize
\setlength{\tabcolsep}{3.5pt}
\caption{Communication size and end-to-end latency under different bandwidths.}
\label{tab:communication_latency}
\resizebox{\textwidth}{!}{%
\begin{tabular}{l rrr rrr rrr}
\toprule
& \multicolumn{3}{c}{\textbf{Baseline}}
& \multicolumn{3}{c}{\textbf{SilentWood (without compression)}}
& \multicolumn{3}{c}{\textbf{SilentWood}} \\
\cmidrule(lr){2-4}
\cmidrule(lr){5-7}
\cmidrule(lr){8-10}
\textbf{Dataset}
& \textbf{Comm. Size} & $\mathbf{B=1}$ Gb/s & $\mathbf{100}$ Mb/s
& \textbf{Comm. Size} & $\mathbf{1}$ Gb/s & $\mathbf{100}$ Mb/s
& \textbf{Comm. Size} & $\mathbf{1}$ Gb/s & $\mathbf{100}$ Mb/s \\
\midrule
spam       & 84.8MB & 312.0s  & 314.1s  & 85.7MB & 10.3s & 12.4s & 22.8MB & 8.2s  & 8.9s  \\
steel      & 84.8MB & 49.4s   & 51.2s   & 85.7MB & 4.9s  & 6.9s  & 16.3MB & 4.0s  & 4.3s  \\
cancer     & 84.8MB & 27.5s   & 29.3s   & 85.7MB & 4.6s  & 6.5s  & 12.3MB & 3.5s  & 3.6s  \\
heart      & 84.8MB & 309.6s  & 311.8s  & 85.7MB & 9.3s  & 11.4s & 7.1MB  & 8.1s  & 8.2s  \\
defect     & 84.8MB & 836.5s  & 838.5s  & 85.7MB & 14.3s & 14.3s & 10.0MB & 12.7s & 13.1s \\
bank       & 84.8MB & 1186.1s & 1187.9s & 85.7MB & 19.9s & 22.6s & 17.5MB & 16.1s & 18.3s \\
pendigits  & 84.8MB & 119.9s  & 121.9s  & 85.7MB & 6.4s  & 7.5s  & 8.4MB  & 4.4s  & 4.9s  \\
ailerons   & 84.8MB & 1072.7s & 1074.6s & 85.7MB & 18.6s & 20.7s & 16.5MB & 15.4s & 16.1s \\
creditcard & 84.8MB & 18.4s   & 20.5s   & 85.7MB & 4.3s  & 6.2s  & 12.3MB & 3.4s  & 4.3s  \\
satellite  & 84.8MB & 31.0s   & 33.0s   & 85.7MB & 4.5s  & 6.4s  & 16.3MB & 3.6s  & 3.8s  \\
elevators  & 84.8MB & 1188.4s & 1190.6s & 85.7MB & 17.4s & 19.6s & 8.7MB  & 14.9s & 16.1s \\
telescope  & 84.8MB & 15.8s   & 17.8s   & 85.7MB & 4.2s  & 6.1s  & 5.8MB  & 3.2s  & 3.6s  \\
mfeat      & 84.8MB & 589.3s  & 591.1s  & 85.7MB & 20.4s & 21.6s & 43.8MB & 17.2s & 18.3s \\
\midrule
\textbf{SilentWood's AVG}
& \textbf{84.8MB} & \textbf{442.8s} & \textbf{444.8s}
& \textbf{85.7MB} & \textbf{10.7s} & \textbf{12.5s}
& \textbf{15.2MB} & \textbf{8.8s} & \textbf{9.5s} \\
\textbf{Speedup}
& n/a & 50.3$\times$ & 46.8$\times$
& n/a & 1.2$\times$ & 1.3$\times$
& n/a & -- & -- \\
\bottomrule
\end{tabular}%
}
\end{table}

\subsection{End-to-end Inference Time across Different Bandwidths}
\label{appendix:bandwidth}
\autoref{tab:communication_latency} is an evaluation for end-to-end inference time and communication data size (C-Size) at latency=200ms and bandwidth B=1Gb/s and 100Mb/s (for these measurements we used Intel Ultra 7 255H, 16 Cores). \sysname{}'s (with ciphertext compression) average communication size is smaller than Baseline (15.2MB $<$ 84.8MB), and its average inference speed is 46.8x faster even at the biggest RTT (200ms) and the smaller bandwidth (100Mb/s). Without ciphertext compression, \sysname{}'s average inference speed is still 35.6x faster than Baseline (444.8s $>$ 12.5s) at the network condition. Without ciphertext compression, \sysname{}'s total communication data size is only marginally bigger than Baseline (85.7MB $>$ 84.8MB), because the BCC protocol's exchanged intermediate path results all compactly fit in 1$\sim$2 ciphertexts.

\clearpage                   
\makeatletter
\let\balance@columns\relax   
\makeatother